\documentclass[english,aps,preprint]{iopart}
\usepackage[T1]{fontenc}
\usepackage[latin9]{inputenc}
\usepackage{geometry}
\usepackage{color}
\usepackage{babel}
\usepackage{array}
\usepackage{float}
\usepackage{booktabs}
\usepackage{multirow}
\usepackage{amstext}
\usepackage{amssymb}
\usepackage{graphicx}
\usepackage{setspace}
\usepackage[numbers]{natbib}
\usepackage[unicode=true,
 bookmarks=true,bookmarksnumbered=false,bookmarksopen=false,
 breaklinks=false,pdfborder={0 0 0},pdfborderstyle={},backref=false,colorlinks=false]
 {hyperref}

\makeatletter

\providecommand{\tabularnewline}{\\}

\usepackage{iopams}
\usepackage{setstack}

\@ifundefined{showcaptionsetup}{}{%
 \PassOptionsToPackage{caption=false}{subfig}}
\usepackage{subfig}
\makeatother

\begin{document}
\title{A first-principles study of the electronic, vibrational, and optical
properties of planar SiC quantum dots}
\author{Rupali Jindal, Vaishali Roondhe, and Alok Shukla{*}}
\address{Department of Physics, Indian Institute of Technology Bombay, Powai,
Mumbai 400076, India}
\ead{rupalijindalpu@gmail.com, oshivaishali@gmail.com, shukla@iitb.ac.in}
\begin{abstract}
With the reported synthesis of a fully planar 2D silicon carbide (SiC)
allotrope, the possibilities of its technological applications are
enormous. Recently, several authors have computationally studied the
structures and electronic properties of a variety of novel infinite
periodic SiC monolayers, in addition to the honeycomb one. In this
work, we perform a systematic first-principles investigation of the
geometry, electronic structure, vibrational, and optical absorption
spectra of several finite, but, fully planar structures of SiC, i.e.,
0D quantum dots (QDs). The sizes of the studied structures are in
the 1.20--2.28 nm range, with their computed HOMO(H)-LUMO(L) gaps
ranging from 0.66 eV to 4.09 eV, i.e., from the IR to the UV region
of the spectrum. The H-L gaps in the SiC QDs are larger as compared
to the band gaps of the corresponding monolayers, confirming the quantum
confinement effects. In spite of covalent bonding in the QDs, Mulliken
charge analysis reveals that Si atoms exhibit positive charges, whereas
the C atoms acquire negative charges, due to the different electron
affinities of the two atoms. Furthermore, a strong structure property
relationship is observed with fingerprints both in the vibrational
and optical spectra. The wide range of H-L gaps in different SiC QDs
makes them well-suited for applications in fields such as photocatalysis,
light-emitting diodes, and solar cells. 
\end{abstract}
\maketitle

\section{Introduction}

Silicon carbide (SiC) is an interesting wide bandgap bulk semiconductor
that has been investigated thoroughly\citep{yakimova2007surface,katoh2012radiation,goldberg2001silicon}
because of its numerous applications in fields such as bioimaging,
photovoltaic cells, optical sensors in the UV region, etc. Its inherent
properties like wide band gap, chemical inertness,\citep{goldberg2001silicon,ledoux1988new}
and high thermal conductivity\citep{kowbel2000high,xie2002thermal,goela2006high}
make it suitable for devices operating at high temperature, frequency,
and voltage such as power electronic devices\citep{536818,casady1996status,cooper2002sic}.
In the crystalline form, SiC exists in over 250 polytypes\textcolor{black}{\citep{verma-krishna-sic-book,JEPPS-SiC-polytypes}}
although the most frequently used polytypes for device applications
have cubic (e.g., 3C-SiC), hexagonal (e.g., 2H-SiC, 4H-SiC, 6H-SiC),
and rhombohedral (e.g., 15R-SiC) crystal structures. Interestingly,
each of these polytypes with different electronic, optical, and vibrational
properties has been widely explored both experimentally, and theoretically.\citep{park1994structural,kackell1994electronic,lee1994first,cobet2000optical,devaty1997optical,kildemo2004optical,karch1996pressure,choyke1997physical}
Over and above these properties, it is worth mentioning that SiC is
an environmental friendly material which can be produced economically
as both Si and C are abundant elements on earth.

\textcolor{black}{The optical spectra of all the polytypes of SiC
lie between 2.416 to 3.33 eV, but, n-type doping modifies its optical
bands to 1-3 eV region \cite{limpijumnong1999optical}. However, bulk
SiC polytypes show weak luminescence characteristics because of their
indirect bandgap \citep{devaty1997optical}. In the present times,
there is a continuous effort to miniaturize devices to the nanoscale
with the aim of tailoring the material properties by exploiting the
quantum size effects. It is well-known that the electronic and optical
properties of matter alter considerably when its dimensionality is
reduced, and can also result in enhanced photoluminescence of otherwise
poorly luminescent materials. With the aim of understanding the influence
of reduced dimensionality on its properties, various nanostructures
of SiC such as one-dimensional (1D) nanowires\citep{kiymaz2016controlled,hu2000synthesis,liang2000large,sun2002formation},
1D nanotubes \cite{menon2004structure,alam2008hybrid,taguchi2005preparation,sun2002formation},
nanoribbons \cite{zhang2010synthesis}, and porous and other complex
nanostructures \cite{fan2006low,chen2019one,ponraj2016sic,wei2006growth}
have been synthesized successfully. Furthermore, nanosized SiC compound
clusters, which can be seen as precursors of SiC QDs, were produced
in the gas phase using the laser vaporization technique and deposited
on various substrates by Melinon }\textcolor{black}{\emph{et al.}}\textcolor{black}{\citep{melinon1998nanostructured}
In our group, first-principles density-functional theory (DFT) studies
on SiC nanoribbons of both armchair and zigzag varieties to explore
their band structures and optical properties have been performed previously\citep{alaal2016first,alaal2017half}.}

\textcolor{black}{Strong c}ovalent bonding leads to stable SiC structures
even in reduced dimensions. In bulk, SiC with four covalent bonds
exhibits $sp^{3}$ hybridization forming four $\sigma$ bonds, whereas
SiC monolayer, due to its planar nature displays $sp^{2}$ hybridization
forming three $\sigma$ bonds, while the remaining $p_{z}$ orbital
laterally overlaps with the neighbouring $p_{z}$ orbitals, forming
a $\pi$-electron band. It has been difficult to isolate graphene-like
monolayer of SiC because in the bulk form, there is no graphitic phase
of SiC held together by Van der Waals forces. All the bulk polytypes
of SiC have strong covalent bonds because of which it is difficult
to isolate a monolayer by exfoliation methods. However, in remarkable
recent works, Chabi \emph{et al.}\citep{chabi2021creation} reported
the synthesis of a fully planar graphene-like honeycomb monolayer
of SiC by employing a wet exfoliation method,\textcolor{black}{{} while
Polley}\textcolor{black}{\emph{ et al.}}\textcolor{black}{\citep{polley2023bottom}
synthesized it on a substrate using a bottom-up approach.}

\textcolor{black}{Now that both 1D (nanotubes, nanowires, etc.) and
2D periodic structures of SiC have been synthesized, one wonders if
its finite 0D structures, i.e., quantum dots (QDs) can also be synthesized
in a controlled manner with tailor-made properties. Because of their
0D (finite) nature, QDs have numerous fascinating properties, as compared
to the higher-dimensional structures made of the same material. These
include edge effects, tunable energy gaps, efficient photoluminescence,
and strong absorption so that their practical implications can be
seen all around, such as in solar cells \citep{raffaelle2002quantum,rahman2021cadmium,mahalingam2021functionalized,shen2008assembly},
electronic devices\citep{yan2019recent,skolnick2004self,nann2011quantum},
light emitting diodes (LEDs) \citep{sun2007bright,dai2014solution,liu2020micro,moon2019stability,shi2017high,yao2016efficient,park2001band},
energy storage and conversion\citep{veeramani2019quantum}, etc. Two
recent works involving the synthesis of SiC-QDs have been reported:
(a) Cao }\textcolor{black}{\emph{et al}}\textcolor{black}{. \citep{cao2018photoluminescent}
synthesised strictly monolayer SiC-QDs of graphitic structures exhibiting
intense photoluminescence, while (b) Mizuno }\textcolor{black}{\emph{et
al}}\textcolor{black}{.\citep{mizuno2020sic} created SiC-QDs of 3D
structures exhibiting $sp^{3}$ hybridization.}\textbf{\textcolor{black}{{}
}}\textcolor{black}{In addition to the widely studied graphitic SiC
monolayer, in recent years several novel 2D monolayers of SiC have
been proposed based on first-principles calculations\citep{qin2015origin,yang-2d-sic-2015,long2021sic}.
Graphene quantum dots (GQDs) possess remarkable properties such as
biocompatibility, low toxicity, and chemical stability, making them
strong candidates for sensing, drug delivery, energy storage and conversion
applications\citep{yang2020revealing,liu2020graphene,sharma2022role,sharma2020four}.
Similarly, SiC QDs find applications in bio-imaging, drug delivery,
dental implants, etc.\citep{sic-qd-n-jrn-chem-2020} Nevertheless,
in the area of optoelectronics SiC-QDs offer several advantages as
compared to GQDs. Some experiments have reported somewhat higher quantum
yield of SiC QDs (7.95\%)\citep{cao2018photoluminescent} as compared
to GQDs (6.9\%)\citep{pan2010hydrothermal}. Additionally, SiC QDs
exhibited extended fluorescence lifetimes of 2.59 $\mu s$\citep{cao2018photoluminescent},
much larger than 7.52 }\textcolor{black}{\emph{ns}}\textcolor{black}{{}
measured for GQDs \citep{ge2014graphene}, and 4.66 $ns$ for MoS$_{2}$
QDs\citep{dai2015tunable}. Higher fluorescence lifetimes of SiC-QDs
will not only facilitate their easier detection but also lead to superior
performance in imaging applications. Furthermore, larger GQDs have
a smaller bandgap, which is unsuitable for higher energy optoelectronics\citep{xi2011honeycomb}.
Although reducing the size of GQDs could potentially address this,
experimental challenges make this approach difficult to implement\citep{xi2011honeycomb,ouarrad2020engineering}.
Because SiC has an inherent band gap due to its hetero-atom character,
the gaps of the SiC QDs will always be larger than the corresponding
GQDs, making them more suitable for UV-Vis optoelectronic applications\citep{benemanskaya2019carbon,xu2018two}.
In the present work, we investigate the existence o}f strictly monolayer
SiC-QDs of a variety of structures, in addition to the graphitic ones,
using a first-principles methodology. Planar SiC QDs, because of their
finite extent, will have reactive dangling bonds on the edges leading
to edge reconstructions resulting in highly asymmetric structures.
Therefore, we have passivated the edges with the H atoms to make them
chemically inert because of which the QDs manage to retain their symmetric
shapes. In previous several works, we have studied the electronic
structure and optical properties of graphene-derived nanostructures
such as graphene nanoribbons as well as quantum dots with H-passivated
edges, employing a Pariser-Parr-Pople (PPP) model based semiempirical
approach\citep{springer-chapter,rai-scientific-reports,basak2021graphene}.
However, in the present work we have employed a DFT-based first-principles
approach, and studied the stability, electronic structure, Raman and
optical absorption spectra of six different strictly planar structures
of SiC-QDs, for which 2D monolayers have been demonstrated to be stable
in the previous works.\citep{qin2015origin,yang-2d-sic-2015,long2021sic}

The remainder of this article is organized as follows. In the next
section, we briefly describe our computational methodology, followed
by a detailed discussion of our results. Finally, we summarize our
work and describe the key conclusions.

\section{Computational Approach}

\begin{figure}[H]
\begin{centering}
\subfloat{\centering{}\includegraphics[scale=0.8]{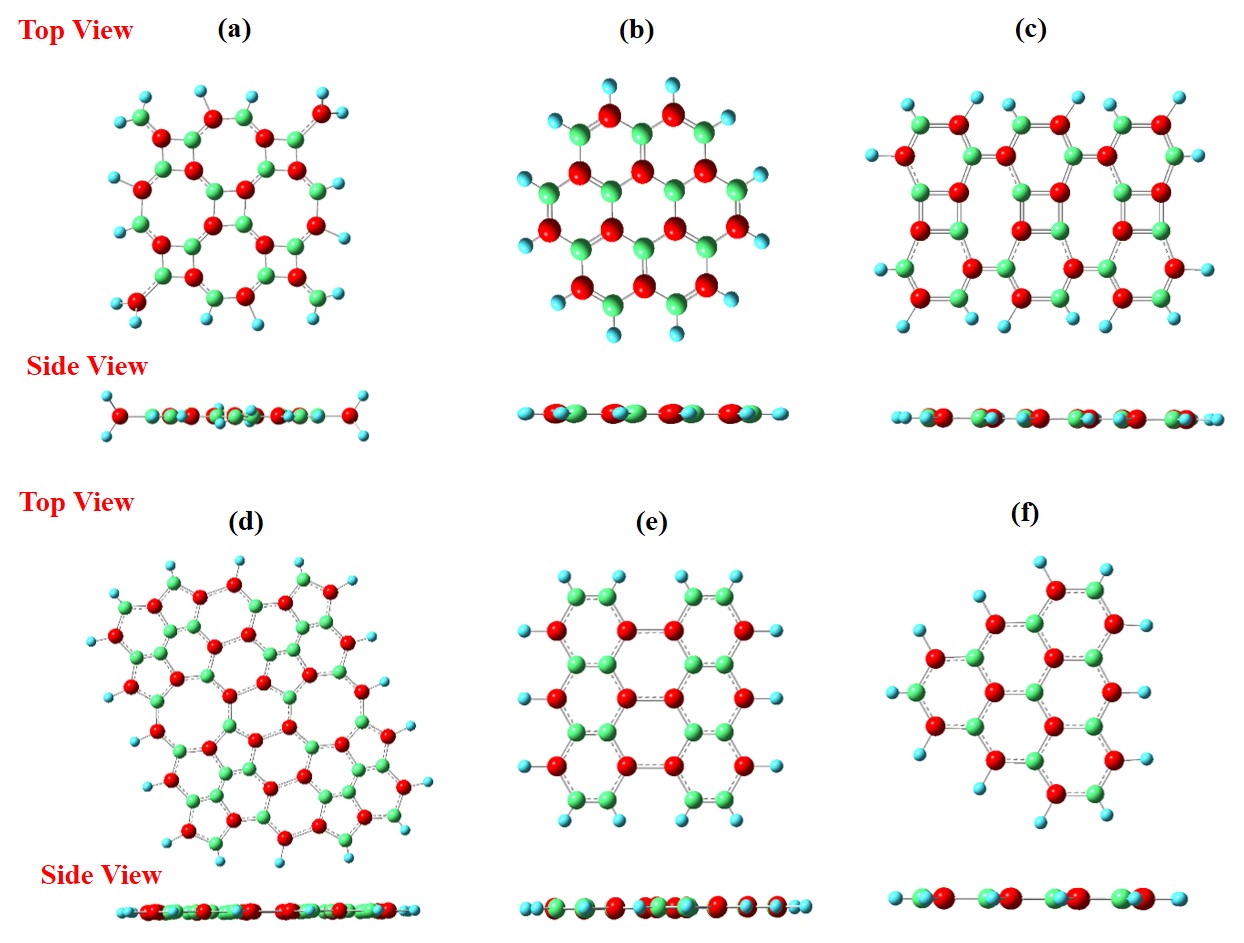}}
\par\end{centering}
\centering{}\caption{\label{fig:input-all} Initial structures of (a) Haeck-SiC (b) Circular-SiC
(c) Tho-SiC (d) Pho-SiC (e) T1-SiC (f) Triangular-SiC, with silicon,
carbon and hydrogen atoms shown in red, green, and blue colors, respectively.}
\end{figure}

In order to investigate the shape and size effect on the properties
of the SiC-QDs six structures are studied (a) Haeck-SiC (b) Circular-SiC
(c) Tho-SiC (d) Pho-SiC (e) T1-SiC (f) Triangular-SiC, where the number
of non-hydrogen atoms is 32, 24, 36, 72, 28, and 22, respectively
(Fig. \ref{fig:input-all}).\textcolor{black}{{} All the initial structures
were taken to be planar and in Figs. \ref{fig:input-all}(b) and \ref{fig:input-all}(f),
the hexagons are arranged in circular and triangular geometrical shapes,
termed Circular-SiC and Triangular-SiC, respectively. In the haeckelite
structure squares are connected with octagons and vice-versa\citep{camacho2015gan,jindal2022density}.
The Tho and Pho structures were initially proposed by Long }\textcolor{black}{\emph{et
al}}\textcolor{black}{.\citep{long2021sic} as 2D-periodic monolayers,
while here we are considering their 0D counterparts, i.e., QDs whose
nomenclature can be explained as follows: (a) The Tho structure is
composed of tetragons (T), hexagons(h), and octagons(o), whereas (b)
the Pho structure contains pentagons(P), hexagons (h), and octagons
(o)\citep{long2021sic}. The structure \ref{fig:input-all}(e) named
T1-SiC was first proposed by Qin et. al. with reference to the linkage
of $\mathrm{Si-Si}$ and $\mathrm{C-C}$ with $\mathrm{Si-C}$ bonds\citep{qin2015origin}.
Of the six QDs considered here, the Circular-SiC and Triangular-SiC
have graphitic structures and are SiC counterparts of hydrocarbon
molecules coronene, and triangulene, respectively. For all the structures
considered here, calculations were performed using the Gaussian16
suite of packages \citep{frisch2009gaussian}, employing a first-principles
density functional th}eory (DFT) \citep{hohenberg1964inhomogeneous,PhysRev.140.A1133,RevModPhys.61.689}
based approac\textcolor{black}{h. Spin-polarized calculations were
performed to investigate the possibility of magnetic ground state
for all the considered QDs. The first five SiC QDs were found to be
stable in non-magnetic spin-singlet states, whereas the Triangular-SiC
QD was found to have a spin-triplet ground state. T}he construction
of initial structures and visualization of the considered QDs has
been done using Gaussview6 \cite{dennington2009gaussview} software.
\textcolor{red}{{} }For performing the calculations, the B3LYP hybrid
functional coupled with the cc-pvDz basis was used. This functional
includes contributions from Becke 88 exchange functional\citep{becke1988density},
and the correlational functional proposed by Lee, Yang, and Parr\citep{lee1988development}.
It has been found to perform very well for semiconductors\citep{tomic2008group},
and the QDs considered in this work are semiconducting nanoparticles
for which the B3LYP functional is used quite frequently.\citep{PhysRevLett.87.276402,PhysRevLett.89.196803,niaz2013size}
After performing the geometry optimization, the Raman intensities
are calculated for all the structures. The lack of any imaginary vibrational
frequencies indicates that the predicted structures of the SiC-QDs
are stable. \textcolor{black}{For the Mulliken charge analysis and
the calculation of the total and partial density of states multiwfn
software is utilized\citep{lu2012multiwfn}.} The optimized structures
are used to calculate the optical absorption spectra using the time-dependent
density functional theory (TD-DFT), employing the same basis set and
functional. 

\section{Results and Discussion}

\subsection{Structural Parameters}

\begin{figure}[H]
\centering{}\includegraphics[scale=0.8]{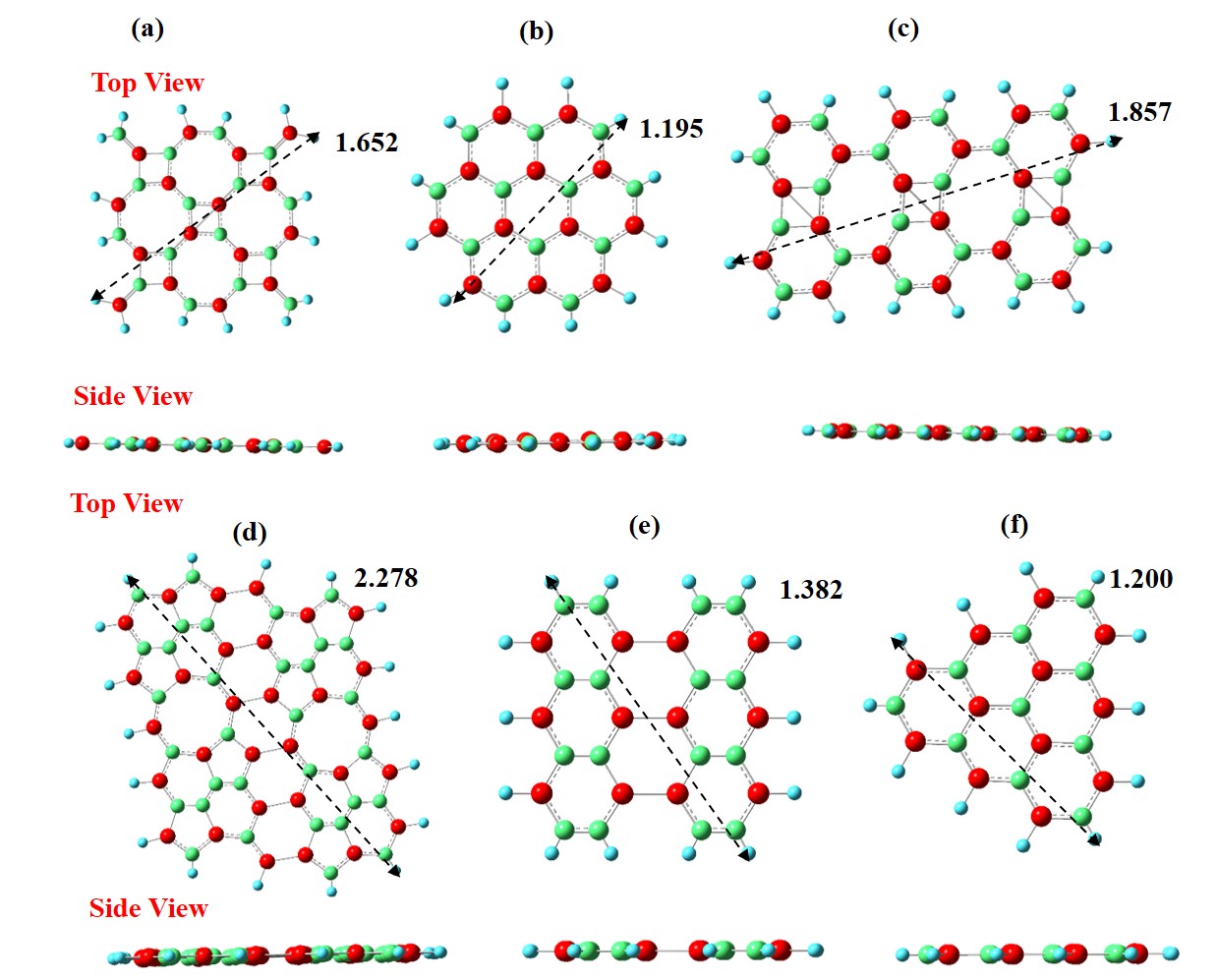}\caption{\label{fig:size-of-SiC}The final optimized structures of (a) Haeck-SiC,
(b) Circular-SiC, (c) Tho-SiC, (d) Pho-SiC (e) T1-SiC, and (f) Triangular-SiC,
along with their sizes in nm. Silicon, carbon and hydrogen atoms are
shown in red, green, and blue colours, respectively.}
\end{figure}

The geometry iterations were performed until the four parameters\textcolor{black}{,
namely maximum force, RMS force, maximum displacement, and RMS displacement
converged for all the QDs indicating stable structures. }In Fig. \ref{fig:size-of-SiC}
we present the final optimized geometries of SiC-QDs considered in
this work. In the figure, we also indicate the sizes of these QDs,
defined as the distances between the farthest hydrogen atoms, and
note that they vary in the range of 1.195-2.278 nm. The detailed structural
data for each QD, i.e., bond lengths and bond angles, are presented
in Tables \ref{tab:bond_lengths-mmav} and \ref{tab:bond-angles-mmav},
respectively. 
\textcolor{black}{{} From the side views of the optimized structures,
it is obvious that each QD is strictly planar, implying that each
atom contributes one $\mathrm{p_{z}}$ electron, i.e., $\pi$ electron,
to the system, while the rest of its three electrons participate in
forming $\mathrm{sp^{2}}$ hybridized $\sigma$ bonds with its neighbours.
The final optimized bond lengths and bond angles show variations from
structure to structure. }

\textcolor{black}{The Haeck-SiC structure contains four squares and
five octagons arranged as shown in Fig. \ref{fig:size-of-SiC}(a),
with Si-C bond lengths varying from $\mathrm{1.716-1.876\ A^{o}}$.
As compared to the initial structure (Fig. \ref{fig:input-all}(a)),
after optimization, we observe only small changes in bond lengths
and angles. The final bond angles $\angle$C-Si-C ($\angle$Si-C-Si)
of the tetragon and octagon rings are in the ranges $91^{o}-95.3^{o}$
and $131.5^{o}-134.5^{o}$ ($85.4^{o}-90.4^{o}$ and $134.4^{o}-138.6^{o}$),
respectively. }

\textcolor{black}{The Circular-SiC comprises seven hexagons with alternative
single and double bonds, as shown in Fig. \ref{fig:size-of-SiC}(b).
The initial $\mathrm{Si-C}$ bond length of Circular-SiC lies between
$\mathrm{1.369-1.425\ A^{o}}$, which increases to $\mathrm{1.792-1.815\ A^{o}}$
after optimization. Also, the $\mathrm{Si-H}$ and $\mathrm{C-H}$
bond length changes from $\mathrm{1.01\ A^{o}}$ to $\mathrm{1.493\ A^{o}}$
and $\mathrm{1.01\ A^{o}}$ to $\mathrm{1.094\ A^{o}}$, respectively.
The bond angles $\angle$Si-C-Si, and $\angle$C-Si-C of Circular-SiC
lie between $118.4^{o}-123.2^{o}$ and $118.8^{o}-122.5^{o}$, as
listed in Table \ref{tab:bond-angles-mmav}.}

\textcolor{black}{Next, the Tho-SiC QD is composed of six hexagons,
three squares, and two octagons arranged, as shown in Fig. \ref{fig:size-of-SiC}(c).
Its minimum and maximum $\mathrm{Si-C}$ bond lengths initially are
$\mathrm{1.456\ A^{o}}$, $\mathrm{1.810\ A^{o}}$, while after optimization
they get elongated to $\mathrm{1.751\ A^{o}}$, $\mathrm{1.835\ A^{o}}$,
respectively. The range of $\angle$C-Si-C ($\angle$Si-C-Si) in tetragon,
hexagon, and octagon is $94.2^{o}-95.1^{o}$, $115.1^{o}-125.4^{o}$,
and $122.8^{o}-142.3^{o}$ ($84.9^{o}-85.8^{o}$, $112^{o}-122^{o}$,
and $122.2^{o}-152.9^{o}$), respectively. }

\textcolor{black}{As seen in Fig. \ref{fig:size-of-SiC}(d), the Pho-SiC
QD is made up of eight pentagons, 13 hexagons, and two octagons. The
initial minimum and maximum $\mathrm{Si-C}$ bond lengths in this
case are $\mathrm{1.752\ A^{o}}$ and $\mathrm{1.880\ A^{o}}$, respectively.
The Pho structure also contains $\mathrm{Si-Si}$ bonds whose initial
bond length varies in between $\mathrm{2.237-2.281\ A^{o}}$. The
corresponding optimized bond lengths of $\mathrm{Si-C}$ and $\mathrm{Si-Si}$
lie between $\mathrm{1.417-2.271\ A^{o}}$ and $\mathrm{2.237-2.268\ A^{o}}$.
Thus, as compared to the initial values, we observe some bond compression
in this case. The optimized bond angles $\angle$C-Si-C ($\angle$Si-C-Si)
of the pentagon, hexagon, and octagon vary between $104.8^{o}-109.3^{o}$
($101^{o}-104.7^{o}$), $115.7^{o}-121.2^{o}$ ($107.5^{o}-122.5^{o}$),
and $131.5^{o}-135.6^{o}$ ($131.3^{o}-140.2^{o}$), respectively. }

\textcolor{black}{Figure \ref{fig:size-of-SiC}(e) shows that T1-SiC
QD has two types of hexagons with four C atoms and two Si atoms in
the first type, and two Si atoms and four C atoms in the second type.
All the initial $\mathrm{Si-C}$, $\mathrm{C-C}$ and $\mathrm{Si-Si}$
bonds were taken to be of same lengths $\mathrm{1.756\ A^{o}}$ ,$\mathrm{1.401\ A^{o}}$,
and $\mathrm{2.219\ A^{o}}$, respectively. After optimization, however,
we note that $\mathrm{C-C}$, $\mathrm{C-Si}$ and $\mathrm{Si-Si}$
bond lengths vary between $\mathrm{1.406-1.463\ A^{o}}$ ,$\mathrm{1.781\ A^{o}-1.831\ A^{o}}$,and
$\mathrm{2.267-2.268\ A^{o}}$, respectively. The optimized $\angle$C-Si-C
($\angle$Si-C-Si) bond angles vary in the range $116.4^{o}-119.9^{o}$
($118.6^{o}-119.3^{o}$), while for all the other bond angles the
range is provided in Table \ref{tab:bond-angles-mmav}. }

\textcolor{black}{Finally, the Triangular-SiC QD is made up of six
hexagons arranged in a triangular fashion with zigzag edges as shown
in Fig. \ref{fig:size-of-SiC}(f.) This is a graphitic structure,
with the optimized C-Si bond lengths in the range $\mathrm{1.770\ A^{o}-1.821\ A^{o}}$,
and the $\angle$C-Si-C ($\angle$Si-C-Si) bond angles in the range
$119.6^{o}-123.3^{o}$ ($118.1^{o}-122.7^{o}$). }

\textcolor{black}{In general, the Si-H and C-H bond lengths show less
variation for all the QDs whose values are recorded in Table \ref{tab:bond_lengths-mmav}.
In summary, the order of Si-C maximum bond length is Pho-SiC > Haeck-SiC
> Tho-SiC > T1-SiC > Triangular-SiC > Circular-SiC. The alteration
in the bond lengths and bond angles of the considered structures suggests
distortions in the shapes of tetragons, hexagons, and octagons, as
compared to their ideal structures. }
\begin{table}[H]
\centering{}\caption{\label{tab:bond_lengths-mmav}Size, formation energies, and bond lengths
of (a) Haeck-SiC (b) Circular-SiC (c) Tho-SiC (d) Pho-SiC (e) T1-SiC
(f) Triangular-SiC}
\begin{tabular}{>{\centering}p{1.5cm}>{\centering}p{2.5cm}>{\centering}p{1.5cm}>{\centering}p{1.8cm}>{\centering}p{1.8cm}>{\centering}p{1.8cm}>{\centering}p{1.8cm}>{\centering}p{1.8cm}}
\toprule 
\multirow{2}{1.5cm}{{\footnotesize{}}%
\begin{tabular}{c}
{\footnotesize{}Structure}\tabularnewline
\end{tabular}} & \multirow{2}{2.5cm}{{\footnotesize{}}%
\begin{tabular}{c}
{\footnotesize{}Size of QD (nm)}\tabularnewline
\end{tabular}} & \multirow{2}{1.5cm}{{\footnotesize{}}%
\begin{tabular}{c}
{\footnotesize{}$E_{b\:}(eV)$}\tabularnewline
\end{tabular}} & \multicolumn{1}{c}{{\footnotesize{}}%
\begin{tabular}{c}
{\footnotesize{}$\mathrm{C-C}$($\text{Å}$)}\tabularnewline
\end{tabular}} & \multicolumn{1}{c}{{\footnotesize{}}%
\begin{tabular}{c}
{\footnotesize{}$\mathrm{Si-Si}$ ($\text{Å}$)}\tabularnewline
\end{tabular}} & \multicolumn{1}{c}{{\footnotesize{}}%
\begin{tabular}{c}
{\footnotesize{}$\mathrm{Si-H}$ ($\text{Å}$)}\tabularnewline
\end{tabular}} & \multicolumn{1}{c}{{\footnotesize{}}%
\begin{tabular}{c}
{\footnotesize{}$\mathrm{C-H}$ ($\text{Å}$)}\tabularnewline
\end{tabular}} & \multicolumn{1}{c}{{\footnotesize{}}%
\begin{tabular}{c}
{\footnotesize{}$\mathrm{Si-C}$ ($\text{Å}$)}\tabularnewline
\end{tabular}}\tabularnewline
\cmidrule{4-8} \cmidrule{5-8} \cmidrule{6-8} \cmidrule{7-8} \cmidrule{8-8} 
 &  &  & {\footnotesize{}}%
\begin{tabular}{c}
{\footnotesize{}Min.-Max.}\tabularnewline
\end{tabular} & {\footnotesize{}}%
\begin{tabular}{c}
{\footnotesize{}Min.-Max.}\tabularnewline
\end{tabular} & {\footnotesize{}}%
\begin{tabular}{c}
{\footnotesize{}Min.-Max.}\tabularnewline
\end{tabular} & {\footnotesize{}}%
\begin{tabular}{c}
{\footnotesize{}Min.-Max.}\tabularnewline
\end{tabular} & {\footnotesize{}}%
\begin{tabular}{c}
{\footnotesize{}Min.-Max.}\tabularnewline
\end{tabular}\tabularnewline
\midrule
{\footnotesize{}}%
\begin{tabular}{c}
{\footnotesize{}Haeck}\tabularnewline
\end{tabular} & {\footnotesize{}}%
\begin{tabular}{c}
{\footnotesize{}1.652}\tabularnewline
\end{tabular} & {\footnotesize{}-5.123} & {\footnotesize{}}%
\begin{tabular}{c}
{\footnotesize{}-}\tabularnewline
\end{tabular} & {\footnotesize{}}%
\begin{tabular}{c}
{\footnotesize{}2.457}\tabularnewline
\end{tabular} & {\footnotesize{}}%
\begin{tabular}{c}
{\footnotesize{}1.488-1.493}\tabularnewline
\end{tabular} & {\footnotesize{}}%
\begin{tabular}{c}
{\footnotesize{}1.093-1.099}\tabularnewline
\end{tabular} & {\footnotesize{}}%
\begin{tabular}{c}
{\footnotesize{}1.716-1.876}\tabularnewline
\end{tabular}\tabularnewline
{\footnotesize{}}%
\begin{tabular}{c}
{\footnotesize{}Circular}\tabularnewline
\end{tabular} & {\footnotesize{}}%
\begin{tabular}{c}
{\footnotesize{}1.195}\tabularnewline
\end{tabular} & {\footnotesize{}-5.281} & {\footnotesize{}}%
\begin{tabular}{c}
{\footnotesize{}-}\tabularnewline
\end{tabular} & {\footnotesize{}}%
\begin{tabular}{c}
{\footnotesize{}-}\tabularnewline
\end{tabular} & {\footnotesize{}}%
\begin{tabular}{c}
{\footnotesize{}1.493}\tabularnewline
\end{tabular} & {\footnotesize{}}%
\begin{tabular}{c}
{\footnotesize{}1.094}\tabularnewline
\end{tabular} & {\footnotesize{}}%
\begin{tabular}{c}
{\footnotesize{}1.792-1.815}\tabularnewline
\end{tabular}\tabularnewline
{\footnotesize{}}%
\begin{tabular}{c}
{\footnotesize{}Tho}\tabularnewline
\end{tabular} & {\footnotesize{}}%
\begin{tabular}{c}
{\footnotesize{}1.857}\tabularnewline
\end{tabular} & {\footnotesize{}-5.306} & {\footnotesize{}}%
\begin{tabular}{c}
{\footnotesize{}-}\tabularnewline
\end{tabular} & {\footnotesize{}}%
\begin{tabular}{c}
{\footnotesize{}2.463-2.477}\tabularnewline
\end{tabular} & {\footnotesize{}}%
\begin{tabular}{c}
{\footnotesize{}1.491-1.493}\tabularnewline
\end{tabular} & {\footnotesize{}}%
\begin{tabular}{c}
{\footnotesize{}1.092-1.094}\tabularnewline
\end{tabular} & {\footnotesize{}}%
\begin{tabular}{c}
{\footnotesize{}1.751-1.835}\tabularnewline
\end{tabular}\tabularnewline
{\footnotesize{}}%
\begin{tabular}{c}
{\footnotesize{}Pho}\tabularnewline
\end{tabular} & {\footnotesize{}}%
\begin{tabular}{c}
{\footnotesize{}2.278}\tabularnewline
\end{tabular} & {\footnotesize{}-5.684} & {\footnotesize{}}%
\begin{tabular}{c}
{\footnotesize{}1.417-1.463}\tabularnewline
\end{tabular} & {\footnotesize{}}%
\begin{tabular}{c}
{\footnotesize{}2.240-2.271}\tabularnewline
\end{tabular} & {\footnotesize{}}%
\begin{tabular}{c}
{\footnotesize{}1.486-1.493}\tabularnewline
\end{tabular} & {\footnotesize{}}%
\begin{tabular}{c}
{\footnotesize{}1.090-1.093}\tabularnewline
\end{tabular} & {\footnotesize{}}%
\begin{tabular}{c}
{\footnotesize{}1.752-1.880}\tabularnewline
\end{tabular}\tabularnewline
{\footnotesize{}}%
\begin{tabular}{c}
{\footnotesize{}T1}\tabularnewline
\end{tabular} & {\footnotesize{}}%
\begin{tabular}{c}
{\footnotesize{}1.382}\tabularnewline
\end{tabular} & {\footnotesize{}-5.357} & {\footnotesize{}}%
\begin{tabular}{c}
{\footnotesize{}1.406-1.463}\tabularnewline
\end{tabular} & {\footnotesize{}}%
\begin{tabular}{c}
{\footnotesize{}2.267-2.268}\tabularnewline
\end{tabular} & {\footnotesize{}}%
\begin{tabular}{c}
{\footnotesize{}1.484-1.485}\tabularnewline
\end{tabular} & {\footnotesize{}}%
\begin{tabular}{c}
{\footnotesize{}1.097}\tabularnewline
\end{tabular} & {\footnotesize{}}%
\begin{tabular}{c}
{\footnotesize{}1.781-1.831}\tabularnewline
\end{tabular}\tabularnewline
{\footnotesize{}}%
\begin{tabular}{c}
{\footnotesize{}Triangular}\tabularnewline
\end{tabular} & {\footnotesize{}}%
\begin{tabular}{c}
{\footnotesize{}1.2}\tabularnewline
\end{tabular} & {\footnotesize{}-4.999} & {\footnotesize{}}%
\begin{tabular}{c}
{\footnotesize{}-}\tabularnewline
\end{tabular} & {\footnotesize{}}%
\begin{tabular}{c}
{\footnotesize{}-}\tabularnewline
\end{tabular} & {\footnotesize{}}%
\begin{tabular}{c}
{\footnotesize{}1.489-1.490}\tabularnewline
\end{tabular} & {\footnotesize{}}%
\begin{tabular}{c}
{\footnotesize{}1.094}\tabularnewline
\end{tabular} & {\footnotesize{}}%
\begin{tabular}{c}
{\footnotesize{}1.770-1.821}\tabularnewline
\end{tabular}\tabularnewline
 &  &  &  &  &  &  & \tabularnewline
\bottomrule
\end{tabular}
\end{table}

\begin{table}[H]
\centering{}\caption{\label{tab:bond-angles-mmav}The minimum, maximum, and average values
of Bond angles}
\begin{tabular}{>{\centering}p{1.1cm}>{\centering}p{1.1cm}>{\centering}p{1.17cm}>{\centering}p{1.17cm}>{\centering}p{1.17cm}>{\centering}p{1.17cm}>{\centering}p{1.17cm}>{\centering}p{1.17cm}>{\centering}p{1.17cm}>{\centering}p{1.17cm}>{\centering}p{1.17cm}>{\centering}p{1.17cm}}
\toprule 
\multirow{2}{1.1cm}{{\scriptsize{}}%
\begin{tabular}{c}
{\scriptsize{}Structure}\tabularnewline
\end{tabular}} & \multirow{2}{1.1cm}{{\scriptsize{}}%
\begin{tabular}{c}
{\scriptsize{}$\mathrm{Shape}$}\tabularnewline
\end{tabular}} & \multicolumn{2}{c}{{\scriptsize{}}%
\begin{tabular}{c}
{\scriptsize{}$\mathrm{\angle C-Si-C}$($^{o}$)}\tabularnewline
\end{tabular}} & \multicolumn{2}{c}{{\scriptsize{}}%
\begin{tabular}{c}
{\scriptsize{}$\mathrm{\angle Si-C-Si}$ ($^{o}$)}\tabularnewline
\end{tabular}} & \multicolumn{2}{c}{{\scriptsize{}}%
\begin{tabular}{c}
{\scriptsize{}$\mathrm{\angle C-Si-Si}$ ($^{o}$)}\tabularnewline
\end{tabular}} & \multicolumn{2}{c}{{\scriptsize{}}%
\begin{tabular}{c}
{\scriptsize{}$\mathrm{\angle C-C-Si}$ ($^{o}$)}\tabularnewline
\end{tabular}} & \multicolumn{2}{c}{{\scriptsize{}}%
\begin{tabular}{c}
{\scriptsize{}$\mathrm{\angle C-C-C}$ ($^{o}$)}\tabularnewline
\end{tabular}}\tabularnewline
\cmidrule{3-12} \cmidrule{4-12} \cmidrule{5-12} \cmidrule{6-12} \cmidrule{7-12} \cmidrule{8-12} \cmidrule{9-12} \cmidrule{10-12} \cmidrule{11-12} \cmidrule{12-12} 
 &  & {\scriptsize{}}%
\begin{tabular}{c}
{\scriptsize{}Min.-Max.}\tabularnewline
\end{tabular} & {\scriptsize{}}%
\begin{tabular}{c}
{\scriptsize{}Average}\tabularnewline
\end{tabular} & {\scriptsize{}}%
\begin{tabular}{c}
{\scriptsize{}Min.-Max.}\tabularnewline
\end{tabular} & {\scriptsize{}}%
\begin{tabular}{c}
{\scriptsize{}Average}\tabularnewline
\end{tabular} & {\scriptsize{}}%
\begin{tabular}{c}
{\scriptsize{}Min.-Max.}\tabularnewline
\end{tabular} & {\scriptsize{}}%
\begin{tabular}{c}
{\scriptsize{}Average}\tabularnewline
\end{tabular} & {\scriptsize{}}%
\begin{tabular}{c}
{\scriptsize{}Min.-Max.}\tabularnewline
\end{tabular} & {\scriptsize{}}%
\begin{tabular}{c}
{\scriptsize{}Average}\tabularnewline
\end{tabular} & {\scriptsize{}}%
\begin{tabular}{c}
{\scriptsize{}Min.-Max.}\tabularnewline
\end{tabular} & {\scriptsize{}}%
\begin{tabular}{c}
{\scriptsize{}Average}\tabularnewline
\end{tabular}\tabularnewline
\midrule
{\scriptsize{}}%
\begin{tabular}{c}
{\scriptsize{}Haeck}\tabularnewline
\end{tabular} & {\scriptsize{}Tetragon} & {\scriptsize{}}%
\begin{tabular}{c}
{\scriptsize{}91-95.3}\tabularnewline
\end{tabular} & {\scriptsize{}}%
\begin{tabular}{c}
{\scriptsize{}93.2}\tabularnewline
\end{tabular} & {\scriptsize{}}%
\begin{tabular}{c}
{\scriptsize{}85.4-90.4}\tabularnewline
\end{tabular} & {\scriptsize{}}%
\begin{tabular}{c}
{\scriptsize{}87.1}\tabularnewline
\end{tabular} & {\scriptsize{}}%
\begin{tabular}{c}
{\scriptsize{}-}\tabularnewline
\end{tabular} & {\scriptsize{}}%
\begin{tabular}{c}
{\scriptsize{}-}\tabularnewline
\end{tabular} & {\scriptsize{}}%
\begin{tabular}{c}
{\scriptsize{}-}\tabularnewline
\end{tabular} & {\scriptsize{}}%
\begin{tabular}{c}
{\scriptsize{}-}\tabularnewline
\end{tabular} & {\scriptsize{}}%
\begin{tabular}{c}
{\scriptsize{}-}\tabularnewline
\end{tabular} & {\scriptsize{}}%
\begin{tabular}{c}
{\scriptsize{}-}\tabularnewline
\end{tabular}\tabularnewline
 & {\scriptsize{}Octagon} & {\scriptsize{}}%
\begin{tabular}{c}
{\scriptsize{}131.5-134.5}\tabularnewline
\end{tabular} & {\scriptsize{}}%
\begin{tabular}{c}
{\scriptsize{}132.8}\tabularnewline
\end{tabular} & {\scriptsize{}}%
\begin{tabular}{c}
{\scriptsize{}134.4-138.6}\tabularnewline
\end{tabular} & {\scriptsize{}}%
\begin{tabular}{c}
{\scriptsize{}136.1}\tabularnewline
\end{tabular} & {\scriptsize{}}%
\begin{tabular}{c}
{\scriptsize{}-}\tabularnewline
\end{tabular} & {\scriptsize{}}%
\begin{tabular}{c}
{\scriptsize{}-}\tabularnewline
\end{tabular} & {\scriptsize{}}%
\begin{tabular}{c}
{\scriptsize{}-}\tabularnewline
\end{tabular} & {\scriptsize{}}%
\begin{tabular}{c}
{\scriptsize{}-}\tabularnewline
\end{tabular} & {\scriptsize{}}%
\begin{tabular}{c}
{\scriptsize{}-}\tabularnewline
\end{tabular} & {\scriptsize{}}%
\begin{tabular}{c}
{\scriptsize{}-}\tabularnewline
\end{tabular}\tabularnewline
{\scriptsize{}}%
\begin{tabular}{c}
{\scriptsize{}Circular}\tabularnewline
\end{tabular} & {\scriptsize{}Hexagon} & {\scriptsize{}}%
\begin{tabular}{c}
{\scriptsize{}118.8-122.5}\tabularnewline
\end{tabular} & {\scriptsize{}}%
\begin{tabular}{c}
{\scriptsize{}120.6}\tabularnewline
\end{tabular} & {\scriptsize{}}%
\begin{tabular}{c}
{\scriptsize{}118.4-123.2}\tabularnewline
\end{tabular} & {\scriptsize{}}%
\begin{tabular}{c}
{\scriptsize{}120.1}\tabularnewline
\end{tabular} & {\scriptsize{}}%
\begin{tabular}{c}
{\scriptsize{}-}\tabularnewline
\end{tabular} & {\scriptsize{}}%
\begin{tabular}{c}
{\scriptsize{}-}\tabularnewline
\end{tabular} & {\scriptsize{}}%
\begin{tabular}{c}
{\scriptsize{}-}\tabularnewline
\end{tabular} & {\scriptsize{}}%
\begin{tabular}{c}
{\scriptsize{}-}\tabularnewline
\end{tabular} & {\scriptsize{}}%
\begin{tabular}{c}
{\scriptsize{}-}\tabularnewline
\end{tabular} & {\scriptsize{}}%
\begin{tabular}{c}
{\scriptsize{}-}\tabularnewline
\end{tabular}\tabularnewline
{\scriptsize{}}%
\begin{tabular}{c}
{\scriptsize{}Tho}\tabularnewline
\end{tabular} & {\scriptsize{}Tetragon} & {\scriptsize{}}%
\begin{tabular}{c}
{\scriptsize{}94.2-95.1}\tabularnewline
\end{tabular} & {\scriptsize{}}%
\begin{tabular}{c}
{\scriptsize{}94.8}\tabularnewline
\end{tabular} & {\scriptsize{}}%
\begin{tabular}{c}
{\scriptsize{}84.9-85.8}\tabularnewline
\end{tabular} & {\scriptsize{}}%
\begin{tabular}{c}
{\scriptsize{}85.2}\tabularnewline
\end{tabular} & {\scriptsize{}}%
\begin{tabular}{c}
{\scriptsize{}-}\tabularnewline
\end{tabular} & {\scriptsize{}}%
\begin{tabular}{c}
{\scriptsize{}-}\tabularnewline
\end{tabular} & {\scriptsize{}}%
\begin{tabular}{c}
{\scriptsize{}-}\tabularnewline
\end{tabular} & {\scriptsize{}}%
\begin{tabular}{c}
{\scriptsize{}-}\tabularnewline
\end{tabular} & {\scriptsize{}}%
\begin{tabular}{c}
{\scriptsize{}-}\tabularnewline
\end{tabular} & {\scriptsize{}}%
\begin{tabular}{c}
{\scriptsize{}-}\tabularnewline
\end{tabular}\tabularnewline
 & {\scriptsize{}Hexagon} & {\scriptsize{}}%
\begin{tabular}{c}
{\scriptsize{}115.1-125.4}\tabularnewline
\end{tabular} & {\scriptsize{}}%
\begin{tabular}{c}
{\scriptsize{}121.3}\tabularnewline
\end{tabular} & {\scriptsize{}}%
\begin{tabular}{c}
{\scriptsize{}112-122}\tabularnewline
\end{tabular} & {\scriptsize{}}%
\begin{tabular}{c}
{\scriptsize{}118.1}\tabularnewline
\end{tabular} & {\scriptsize{}}%
\begin{tabular}{c}
{\scriptsize{}-}\tabularnewline
\end{tabular} & {\scriptsize{}}%
\begin{tabular}{c}
{\scriptsize{}-}\tabularnewline
\end{tabular} & {\scriptsize{}}%
\begin{tabular}{c}
{\scriptsize{}-}\tabularnewline
\end{tabular} & {\scriptsize{}}%
\begin{tabular}{c}
{\scriptsize{}-}\tabularnewline
\end{tabular} & {\scriptsize{}}%
\begin{tabular}{c}
{\scriptsize{}-}\tabularnewline
\end{tabular} & {\scriptsize{}}%
\begin{tabular}{c}
{\scriptsize{}-}\tabularnewline
\end{tabular}\tabularnewline
 & {\scriptsize{}Octagon} & {\scriptsize{}}%
\begin{tabular}{c}
{\scriptsize{}122.8-142.3}\tabularnewline
\end{tabular} & {\scriptsize{}}%
\begin{tabular}{c}
{\scriptsize{}132.4}\tabularnewline
\end{tabular} & {\scriptsize{}}%
\begin{tabular}{c}
{\scriptsize{}122.2-152.9}\tabularnewline
\end{tabular} & {\scriptsize{}}%
\begin{tabular}{c}
{\scriptsize{}137.6}\tabularnewline
\end{tabular} & {\scriptsize{}}%
\begin{tabular}{c}
{\scriptsize{}-}\tabularnewline
\end{tabular} & {\scriptsize{}}%
\begin{tabular}{c}
{\scriptsize{}-}\tabularnewline
\end{tabular} & {\scriptsize{}}%
\begin{tabular}{c}
{\scriptsize{}-}\tabularnewline
\end{tabular} & {\scriptsize{}}%
\begin{tabular}{c}
{\scriptsize{}-}\tabularnewline
\end{tabular} & {\scriptsize{}}%
\begin{tabular}{c}
{\scriptsize{}-}\tabularnewline
\end{tabular} & {\scriptsize{}}%
\begin{tabular}{c}
{\scriptsize{}-}\tabularnewline
\end{tabular}\tabularnewline
{\scriptsize{}}%
\begin{tabular}{c}
{\scriptsize{}Pho}\tabularnewline
\end{tabular} & {\scriptsize{}Pentagon} & {\scriptsize{}}%
\begin{tabular}{c}
{\scriptsize{}104.8-109.3}\tabularnewline
\end{tabular} & {\scriptsize{}}%
\begin{tabular}{c}
{\scriptsize{}106.8}\tabularnewline
\end{tabular} & {\scriptsize{}}%
\begin{tabular}{c}
{\scriptsize{}101-104.7}\tabularnewline
\end{tabular} & {\scriptsize{}}%
\begin{tabular}{c}
{\scriptsize{}102.8}\tabularnewline
\end{tabular} & {\scriptsize{}}%
\begin{tabular}{c}
{\scriptsize{}-}\tabularnewline
\end{tabular} & {\scriptsize{}}%
\begin{tabular}{c}
{\scriptsize{}-}\tabularnewline
\end{tabular} & {\scriptsize{}}%
\begin{tabular}{c}
{\scriptsize{}110.2-113.8}\tabularnewline
\end{tabular} & {\scriptsize{}}%
\begin{tabular}{c}
{\scriptsize{}111.8}\tabularnewline
\end{tabular} & {\scriptsize{}}%
\begin{tabular}{c}
{\scriptsize{}-}\tabularnewline
\end{tabular} & {\scriptsize{}}%
\begin{tabular}{c}
{\scriptsize{}-}\tabularnewline
\end{tabular}\tabularnewline
 & {\scriptsize{}Hexagon} & {\scriptsize{}}%
\begin{tabular}{c}
{\scriptsize{}115.7-121.2}\tabularnewline
\end{tabular} & {\scriptsize{}}%
\begin{tabular}{c}
{\scriptsize{}119}\tabularnewline
\end{tabular} & {\scriptsize{}}%
\begin{tabular}{c}
{\scriptsize{}107.5-122.5}\tabularnewline
\end{tabular} & {\scriptsize{}}%
\begin{tabular}{c}
{\scriptsize{}115.7}\tabularnewline
\end{tabular} & {\scriptsize{}}%
\begin{tabular}{c}
{\scriptsize{}112.3-128.8}\tabularnewline
\end{tabular} & {\scriptsize{}}%
\begin{tabular}{c}
{\scriptsize{}120}\tabularnewline
\end{tabular} & {\scriptsize{}}%
\begin{tabular}{c}
{\scriptsize{}122.9-126}\tabularnewline
\end{tabular} & {\scriptsize{}}%
\begin{tabular}{c}
{\scriptsize{}124.1}\tabularnewline
\end{tabular} & {\scriptsize{}}%
\begin{tabular}{c}
{\scriptsize{}123.7-124.9}\tabularnewline
\end{tabular} & {\scriptsize{}}%
\begin{tabular}{c}
{\scriptsize{}124.2}\tabularnewline
\end{tabular}\tabularnewline
 & {\scriptsize{}Octagon} & {\scriptsize{}}%
\begin{tabular}{c}
{\scriptsize{}131.5-135.6}\tabularnewline
\end{tabular} & {\scriptsize{}}%
\begin{tabular}{c}
{\scriptsize{}133.9}\tabularnewline
\end{tabular} & {\scriptsize{}}%
\begin{tabular}{c}
{\scriptsize{}131.3-140.2}\tabularnewline
\end{tabular} & {\scriptsize{}}%
\begin{tabular}{c}
{\scriptsize{}136.0}\tabularnewline
\end{tabular} & {\scriptsize{}}%
\begin{tabular}{c}
{\scriptsize{}-}\tabularnewline
\end{tabular} & {\scriptsize{}}%
\begin{tabular}{c}
{\scriptsize{}-}\tabularnewline
\end{tabular} & {\scriptsize{}}%
\begin{tabular}{c}
{\scriptsize{}-}\tabularnewline
\end{tabular} & {\scriptsize{}}%
\begin{tabular}{c}
{\scriptsize{}-}\tabularnewline
\end{tabular} & {\scriptsize{}}%
\begin{tabular}{c}
{\scriptsize{}-}\tabularnewline
\end{tabular} & {\scriptsize{}}%
\begin{tabular}{c}
{\scriptsize{}-}\tabularnewline
\end{tabular}\tabularnewline
{\scriptsize{}}%
\begin{tabular}{c}
{\scriptsize{}T1}\tabularnewline
\end{tabular} & {\scriptsize{}Hexagon} & {\scriptsize{}}%
\begin{tabular}{c}
{\scriptsize{}116.4-119.9}\tabularnewline
\end{tabular} & {\scriptsize{}}%
\begin{tabular}{c}
{\scriptsize{}117.7}\tabularnewline
\end{tabular} & {\scriptsize{}}%
\begin{tabular}{c}
{\scriptsize{}118.6-119.3}\tabularnewline
\end{tabular} & {\scriptsize{}}%
\begin{tabular}{c}
{\scriptsize{}119}\tabularnewline
\end{tabular} & {\scriptsize{}}%
\begin{tabular}{c}
{\scriptsize{}120.6-123.0}\tabularnewline
\end{tabular} & {\scriptsize{}}%
\begin{tabular}{c}
{\scriptsize{}121.5}\tabularnewline
\end{tabular} & {\scriptsize{}}%
\begin{tabular}{c}
{\scriptsize{}120.2-122.8}\tabularnewline
\end{tabular} & {\scriptsize{}}%
\begin{tabular}{c}
{\scriptsize{}121.1}\tabularnewline
\end{tabular} & {\scriptsize{}}%
\begin{tabular}{c}
{\scriptsize{}-}\tabularnewline
\end{tabular} & {\scriptsize{}}%
\begin{tabular}{c}
{\scriptsize{}-}\tabularnewline
\end{tabular}\tabularnewline
{\scriptsize{}}%
\begin{tabular}{c}
{\scriptsize{}Triangular}\tabularnewline
\end{tabular} & {\scriptsize{}Hexagon} & {\scriptsize{}}%
\begin{tabular}{c}
{\scriptsize{}119.6-123.3}\tabularnewline
\end{tabular} & {\scriptsize{}}%
\begin{tabular}{c}
{\scriptsize{}120.9}\tabularnewline
\end{tabular} & {\scriptsize{}}%
\begin{tabular}{c}
{\scriptsize{}118.1-122.7}\tabularnewline
\end{tabular} & {\scriptsize{}}%
\begin{tabular}{c}
{\scriptsize{}120}\tabularnewline
\end{tabular} & {\scriptsize{}}%
\begin{tabular}{c}
{\scriptsize{}-}\tabularnewline
\end{tabular} & {\scriptsize{}}%
\begin{tabular}{c}
{\scriptsize{}-}\tabularnewline
\end{tabular} & {\scriptsize{}}%
\begin{tabular}{c}
{\scriptsize{}-}\tabularnewline
\end{tabular} & {\scriptsize{}}%
\begin{tabular}{c}
{\scriptsize{}-}\tabularnewline
\end{tabular} & {\scriptsize{}}%
\begin{tabular}{c}
{\scriptsize{}-}\tabularnewline
\end{tabular} & {\scriptsize{}}%
\begin{tabular}{c}
{\scriptsize{}-}\tabularnewline
\end{tabular}\tabularnewline
 &  &  &  &  &  &  &  &  &  &  & \tabularnewline
\bottomrule
\end{tabular}
\end{table}

\subsection{Binding Energies}

To analyze the energetic stability of the SiC QDs, the binding energy
per atom is calculated using the formula 
\[
E_{b}=\frac{1}{N_{atom}}(E_{total}-n_{H}E_{H}-n_{C}E_{C}-n_{Si}E_{Si}),
\]

where $n_{H},\;n_{C},\;n_{Si},\;N_{atom}$ respectively represent
the numbers of H, C, Si atoms in the SiC-QD, while the total number
of atoms $N_{atom}=n_{H}+n_{C}+n_{Si}$. The calculated energy of
the $\mathrm{SiC}$ QD is represented by $E_{total}$ here, whereas
$E_{H}$, $E_{C}$, $E_{N}$ are the energies of the respective isolated
H, C, and Si atoms. The calculated binding energies for all the SiC
QDs are presented in Table \ref{tab:bond_lengths-mmav} and are negative
indicating that they can be synthesized in an energetically favourable
way. The calculated binding energies per atom are close to each other,
but their values in descending order inclusive of the sign follow
Triangular-SiC > Haeck-SiC > Circular-SiC > Tho-SiC > T1-SiC > Pho-SiC.
As the Pho-SiC possesses the most negative binding energy, its formation
is most favourable compared to all other QDs. 

\subsection{Electronic Structure}

\subsubsection{Nature of Molecular Orbitals}

\begin{figure}[H]
\centering{}\subfloat{\centering{}\includegraphics[scale=0.5]{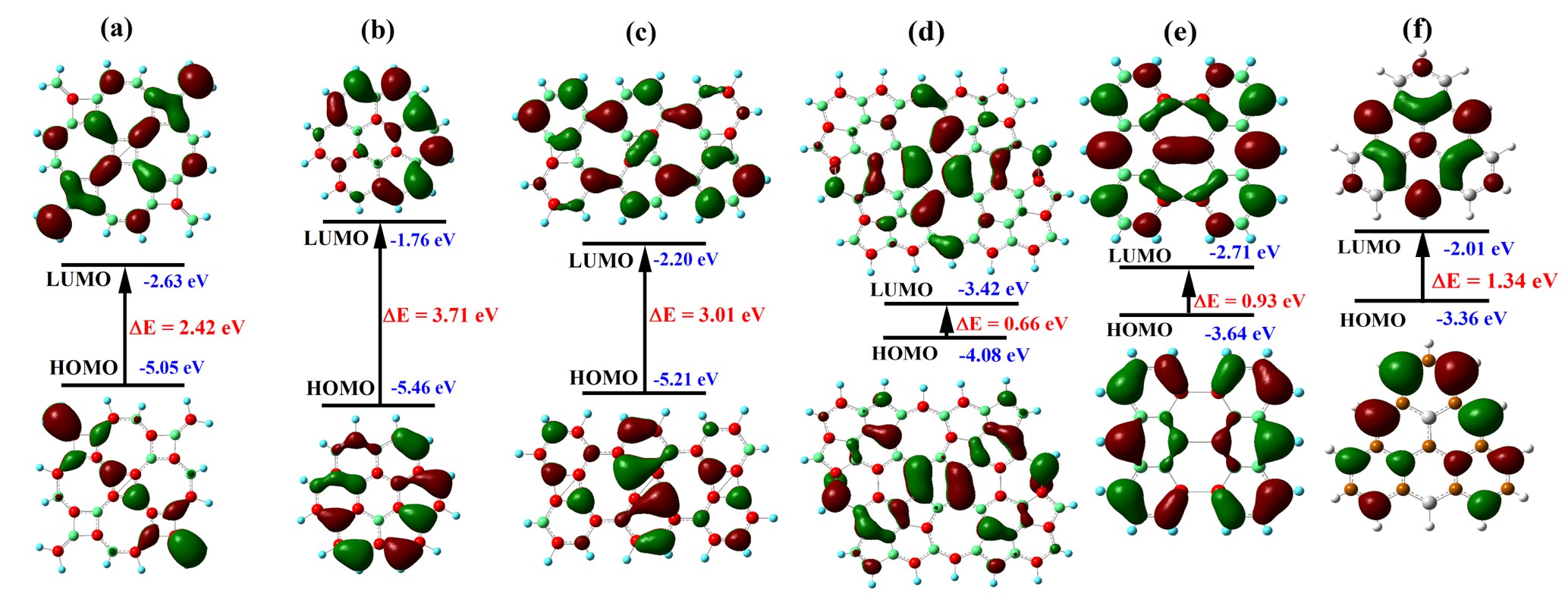}}
\caption{\label{fig:H-L_gap} Plots of HOMO (H) and LUMO (L) orbitals of (a)
Haeck-SiC (b) Circular-SiC (c) Tho-SiC (d) Pho-SiC(e) T1-SiC (f) Triangular-SiC.
The H-L gap of each QD is also indicated.}
\end{figure}

For analyzing the electronic structure and optical properties of SiC
QDs, it is important to understand the nature of their frontier molecular
orbitals (MOs), in general, and HOMO (Highest Occupied Molecular Orbital)
and LUMO (Lowest Unoccupied Molecular Orbital) and their energies,
in particular. \citep{martinez2009local} Because the final optimized
geometries of all the SiC QDs are planar (see Fig. \ref{fig:size-of-SiC}),
we expect the frontier MOs to have the delocalized $\pi$-electron
character. In table \ref{tab:comparison_with_other_work} and Fig.
\ref{fig:H-L_gap} we present the plots of HOMO (H) and LUMO (L) orbitals
of each SiC QD from which it is obvious that they are of $\pi$-type,
delocalized on the Si and C atoms of the QDs, but with a negligible
presence on the passivating hydrogen atoms. In Table \ref{tab:frontier-mo-pi-char}
we list the $\pi/\pi^{*}$ orbitals closest to the Fermi energy, from
which it is obvious that it is not just the H/L orbitals which have
the $\pi$ character, rather a large number of occupied and unoccupied
frontier orbitals have that character. This is similar to the case
of planar aromatic hydrocarbons, therefore, we expect the electronic
properties and the low-lying excited states of these SiC QDs to be
determined by their $\pi$-electron dynamics.

Furthermore, the $H$ and $L$ energy values characterize the electrophilic
and nucleophilic behaviour of the molecules\citep{bradley1973frontier,kiyooka2013intrinsic}.\textcolor{green}{{}
}\textcolor{black}{In Table \ref{tab:comparison_with_other_work}
we compare our calculated H-L gaps of the SiC QDs, to the band gaps
of their periodic monolayer counterparts computed using the VASP code\citep{kresse1993ab,kresse1996efficiency,kresse1996efficient,kresse1999ultrasoft},
we conclude that the gaps increase from 2D to 0D, in accordance with
the quantum confinement effect.} For the calculation of the band gaps
listed in the last column of Table \ref{tab:comparison_with_other_work},
two functionals GGA-PBE and HSE06 (data written as PBE/HSE06 in column
IV) were utilized by Long et al. for 2D SiC siligraphene structures\citep{long2021sic}.
\begin{table}
\centering{}\caption{\label{tab:comparison_with_other_work} Our calculated H-L gaps of
various SiC QDs compared to the band gaps of their 2D-periodic counterparts.
Above $\alpha$ denotes spin up, while $\beta$ denotes spin down
in the UHF calculations. Because of the UHF open-shell triplet ground
state, Triangular-SiC has different H-L gaps for up- and down-spin
electrons.}
\begin{tabular}{ccccc}
\toprule 
QD Structure & $E_{HL}$(eV) & periodic monolayer & $E_{HL}$(eV) & \tabularnewline
\midrule
\midrule 
Haeck-SiC & 2.42 & T-SiC & 2.24/3.07\citep{long2021sic} & \tabularnewline
\midrule 
Circular-SiC & 3.71 & g-SiC & 2.56/3.48\citep{long2021sic} & \tabularnewline
\midrule 
Tho-SiC & 3.01 & Tho-SiC & 1.43/2.16\citep{long2021sic} & \tabularnewline
\midrule 
Pho-SiC & 0.66 & Pho-SiC & Metallic\citep{long2021sic} & \tabularnewline
\midrule 
T1-SiC & 0.93 & T1-SiC & 0.00\citep{long2021sic} & \tabularnewline
\midrule 
Triangular-SiC & 1.34 ($\alpha$)/4.09($\beta$) & g-SiC & \textcolor{black}{2.56/3.48}\citep{long2021sic} & \tabularnewline
\bottomrule
\end{tabular}
\end{table}

\begin{table}[H]
\begin{centering}
\caption{\label{tab:frontier-mo-pi-char}Frontier molecular orbitals with $\pi$
and $\pi^{*}$character of the considered SiC QDs. Our notation implies,
for example, that for Tho-SiC all the orbitals starting from $H-10$
to $H$ and $L$ to $L+9$ have $\pi$ and $\pi^{*}$ characters,
respectively.}
\begin{tabular}{ccccccc}
\toprule 
Structure &  &  & $\pi$ orbital &  &  & $\pi^{*}$ orbital\tabularnewline
\midrule
\midrule 
Haeck-SiC &  &  & $H-10:H$ &  &  & $L:L+6$\tabularnewline
\midrule 
Circular-SiC &  &  & $H-8:H$ &  &  & $L:L+6$\tabularnewline
\midrule 
Tho-SiC &  &  & $H-10:H$ &  &  & $L:L+9$\tabularnewline
\midrule 
Pho-SiC &  &  & $H-14:H$ &  &  & $L:L+11$\tabularnewline
\midrule 
T1-SiC &  &  & $H-5:H$ &  &  & $L:L+5$\tabularnewline
\midrule 
Triangular-SiC &  &  & $H-8:H$ &  &  & $L:L+5$\tabularnewline
\bottomrule
\end{tabular}
\par\end{centering}
\end{table}

\subsubsection{Mulliken Charge Analysis}

The Mulliken charge varies with the atomic arrangement in the molecule\citep{azhagiri2014molecular},
therefore, it provides useful information regarding the structural
dependence of the charge distribution in a molecule, and, thus, on
the nature of the chemical bonding in the system.\citep{jasmine2015vibrational}
Although both Si and C are group IV elements, because of their different
electron affinities and ionization potentials, we expect the SiC structures
to exhibit polar covalent bonds, instead of pure covalent bonds displayed
in pure Si or C based systems. To verify that, we have performed the
Mulliken charge analysis of these SiC QDs and divided it into two
parts: a) Structural comparison, and b) atomic site comparison (electrophilic
or nucleophilic) within a given QD. The charge on each SiC QD is arranged
such that the net charge on the molecule is zero. 

In all the QDs, Si atoms exhibit inhomogeneous positive charges, whereas
C atoms exhibit negative charges. This behaviour can be explained
based on the electronegativity values; C being more electronegative
than Si and H exhibits negative charges, whereas Si and H atoms attached
to C exhibit positive charges. In the Si-H bond, Si is more electropositive,
so H attached to Si attains a negative charge. The total charge gained
by the Si (T$_{\mathrm{Si}}$) and lost by the C (T$_{\mathrm{C}}$)
for all the SiC QDs is presented in Table \ref{tab:Mulliken_all_value}.
The overall zero charge in a given QD can be confirmed by adding the
total charges of H (T$_{\mathrm{H}}$) with T$_{\mathrm{Si}}$ and
T$_{\mathrm{C}}$. The maximum positive and negative charges exhibited
by the SiC QDs, also presented in Table \ref{tab:Mulliken_all_value},
help in identifying the most electrophilic and nucleophilic sites. 

\begin{table}
\centering{}\caption{\label{tab:Mulliken_all_value} The maximum charge on a Si atom, the
maximum charge on a C atom, and the total Mulliken charges carried
by all the C atoms (T$_{\mathrm{C}}$), Si atoms (T$_{\mathrm{Si}}$),
and H atoms (T$_{\mathrm{H}}$) for each SiC QD considered in this
work.}
\begin{tabular}{cccccc}
\toprule 
 & Maximum Charge on Si & Maximum Charge on C & T$_{\mathrm{C}}$ & T$_{\mathrm{Si}}$ & T$_{\mathrm{H}}$\tabularnewline
\midrule
\midrule 
Haeck & 0.369 & -0.333 & -4.316 & 4.237 & 0.079\tabularnewline
Circular & 0.407 & -0.354 & -3.724 & 3.956 & -0.232\tabularnewline
Tho & 0.397 & -0.366 & -5.703 & 6.003 & -0.301\tabularnewline
Pho & 0.399 & -0.359 & -7.576 & 7.953 & -0.377\tabularnewline
T1 & 0.266 & -0.235 & -2.579 & 2.79 & -0.212\tabularnewline
Triangular & 0.438 & -0.369 & -3.055 & 3.411 & -0.356\tabularnewline
\bottomrule
\end{tabular}
\end{table}

\subsubsection{Global Reactivity Descriptors}

The values of other global reactivity descriptors that depend on the
frontier orbital energies, namely, ionization potential ($IP$), electron
affinity ($EA$), chemical potential ($\mu$), chemical hardness ($\eta$),
electrophilicity index ($\omega$), and chemical softness ($S$) \citep{parr1983absolute,pearson1986absolute,pearson1992chemical}
are presented in Table \ref{tab:all_electronic_parameters}. Based
on its largest H-L gap $E_{HL}=3.71$ eV (see Fig. \ref{fig:H-L_gap}),
we predict the circular-SiC QD to possess high kinetic stability and
low chemical reactivity \citep{aihara1999reduced}. Its lowest energy
value (-5.46 eV) of $H$ and high energy value (-1.76 eV) of $L$
as compared to other structures, do not favour charge transfer. Based
on the H-L gap criterion, the QDs can be arranged in the descending
order of kinetic stability as Circular-SiC > Tho-SiC > Haeck-SiC >
Triangular-SiC($\alpha$) >T1-SiC > Pho-SiC. The tendency to oppose
the change in electron distribution is quantified by the chemical
hardness $\eta$ defined as $(IP-EA)/2$, while the ease in polarization
is described by the chemical softness $S$, defined as $S=1/\eta$.
Accordingly, Pho-SiC is the easiest to polarize with the highest softness
value of 3.03, whereas the Circular-SiC is the hardest to polarize
with the least value of $S$ = 0.54.\textcolor{black}{{} The high chemical
potential $\mu=-(E_{H}+E_{L})/2$ facilitates the charge transfer
that leads to high molecular reactivity.} The global electrophilicity
index $\omega=\mu^{2}/2\eta$ defines how much additional electronic
charge the molecule can attain while remaining structurally stable.
According to this, the maximum stabilization will be attained by Pho-SiC
with a large $\omega$ value of 21.31 after electronic charge transfer
between H to L. The dipole moment ($p)$ is the global parameter that
quantifies the net polarization in the ground states of molecules.\citep{pandey2021hybrid}.
From Table \ref{tab:all_electronic_parameters} it is obvious that
the net ground-state dipole moment for the SiC QDs is very small,
i.e., of the order of $10^{-3}$, or smaller. 

\begin{table}
\centering{}\caption{\label{tab:all_electronic_parameters}Calculated global reactivity
parameters for Haeck-SiC, Circular-SiC, Tho-SiC, Pho-SiC, T1-SiC,
and Triangular-SiC QDs.}
\begin{tabular}{c>{\centering}p{1.1cm}>{\centering}p{1.1cm}>{\centering}p{1.1cm}>{\centering}p{1.1cm}>{\centering}p{1.1cm}>{\centering}p{1.1cm}>{\centering}p{1.1cm}>{\centering}p{1.1cm}>{\centering}p{1.1cm}>{\centering}p{1.1cm}}
\toprule 
{\small{}Structure} & {\small{}$E_{H}$(eV)} & {\small{}$E_{L}$(eV)} & {\small{}$E_{HL}$(eV)} & {\small{}$IP$(eV)} & {\small{}$EA$(eV)} & {\small{}$\mu$(eV)} & {\small{}$\eta$(eV)} & {\small{}$\omega$(eV)} & {\small{}$S$($\mathrm{eV^{-1}}$)} & {\small{}$p$(Debye)}\tabularnewline
\midrule
\midrule 
{\small{}Haeck-SiC} & {\small{}-5.05} & {\small{}-2.63} & {\small{}2.42} & {\small{}5.05} & {\small{}2.63} & {\small{}-3.84} & {\small{}1.21} & {\small{}6.09} & {\small{}0.83} & {\small{}0.001}\tabularnewline
\midrule 
{\small{}Circular-SiC} & {\small{}-5.46} & {\small{}-1.76} & {\small{}3.71} & {\small{}5.46} & {\small{}1.76} & {\small{}-3.61} & {\small{}1.85} & {\small{}3.51} & {\small{}0.54} & {\small{}0.001}\tabularnewline
\midrule 
{\small{}Tho-SiC} & {\small{}-5.21} & {\small{}-2.2} & {\small{}3.01} & {\small{}5.21} & {\small{}2.2} & {\small{}-3.71} & {\small{}1.5} & {\small{}4.56} & {\small{}0.66} & {\small{}0.006}\tabularnewline
\midrule 
{\small{}Pho-SiC} & {\small{}-4.08} & {\small{}-3.42} & {\small{}0.66} & {\small{}4.08} & {\small{}3.42} & {\small{}-3.75} & {\small{}0.33} & {\small{}21.31} & {\small{}3.03} & {\small{}0.001}\tabularnewline
\midrule 
{\small{}T1-SiC} & {\small{}-3.64} & {\small{}-2.71} & {\small{}0.93} & {\small{}3.64} & {\small{}2.71} & {\small{}-3.17} & {\small{}0.47} & {\small{}10.84} & {\small{}2.15} & {\small{}0.0001}\tabularnewline
\midrule 
{\small{}Triangular-SiC} & {\small{}-3.36} & {\small{}-2.01} & {\small{}1.34} & {\small{}3.36} & {\small{}2.01} & {\small{}-2.69} & {\small{}0.67} & {\small{}5.37} & {\small{}1.49} & {\small{}0.001}\tabularnewline
\bottomrule
\end{tabular}
\end{table}

\subsection{Raman Spectra}

\begin{figure}[H]
\begin{centering}
\includegraphics[scale=0.6]{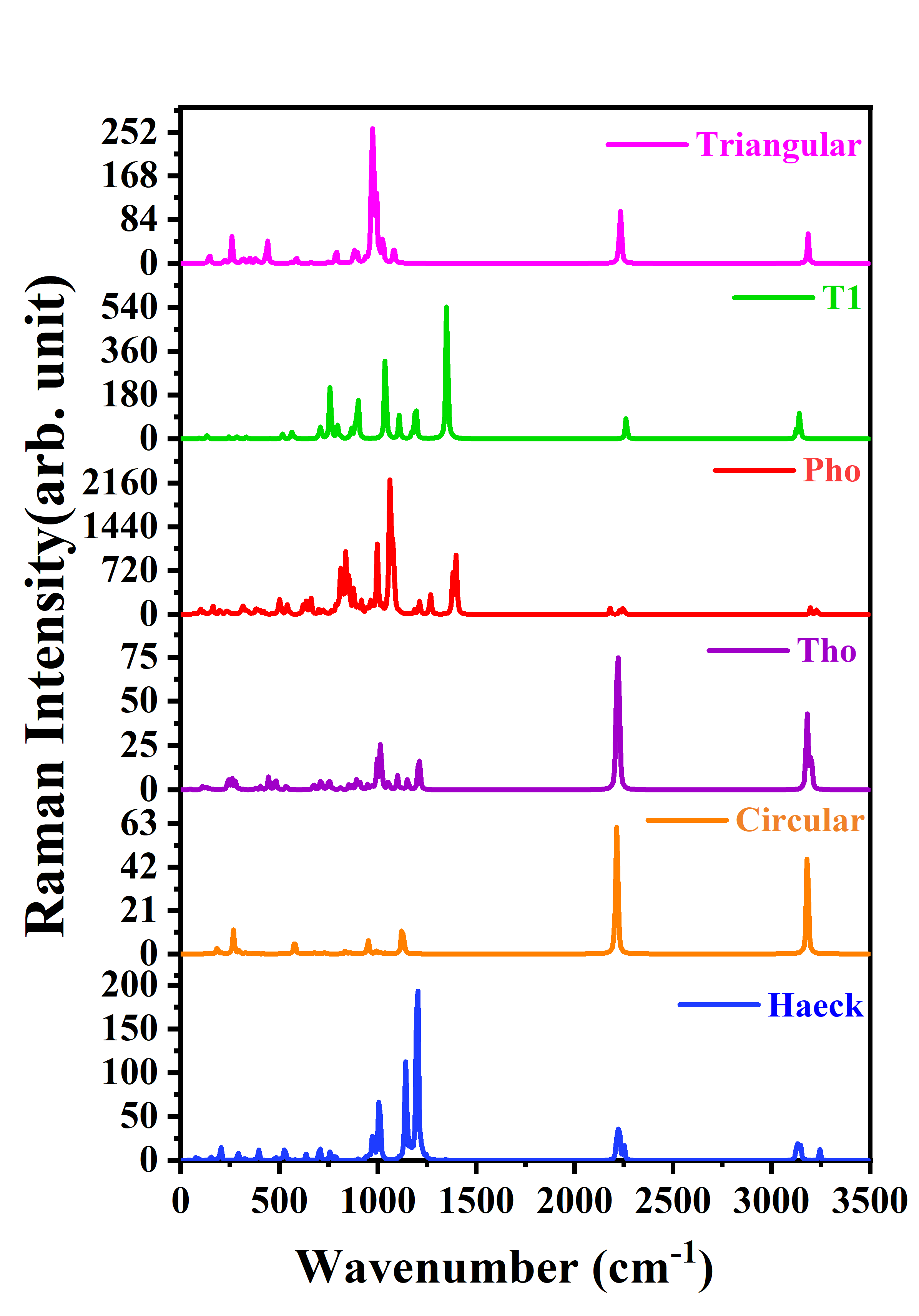}
\par\end{centering}
\centering{}\caption{\label{fig:Raman-all} Calculated Raman spectra of (a) Haeck-SiC,
(b) Circular-SiC, (c) Tho-SiC, (d) Pho-SiC, (e) T1-SiC, and (f) Triangular-SiC,
computed using the B3LYP functional in the frequency range 0-3500
cm$^{-1}$.}
\end{figure}

Our vibrational frequency analysis calculations for the SiC-QDs predicted
all the frequencies to be real, indicating that the structures considered
in this work are dynamically stable. Therefore, we calculated their
Raman spectra presented in Fig. \ref{fig:Raman-all} and discussed
next in detail. 

The Haeck-SiC QD has 48 atoms, i.e., it has $(3\times48)-6=148$ vibrational
modes. The low-frequency region, $0-1010\:cm^{-1}$ corresponds to
$\mathrm{Si-C}$, $\mathrm{C-H}$, and $\mathrm{Si-H}$ in-plane and
out-of-plane bending modes with the strongest intensity at $1004.41\:cm^{-1}$
due to $\mathrm{CH_{2}}$ symmetric bending along with attached $\mathrm{Si-C}$
double bond stretching in the $y$ direction. The second $1010-1300\:cm^{-1}$
frequency region contains in-plane $\mathrm{Si-C}$ stretching, $\mathrm{C-H}$,
and $\mathrm{Si-C}$ bending modes. The highest intensity in this
region is attributed to the single bond $\mathrm{Si-C}$ stretching
along the $x$-axis coupled with in-plane $\mathrm{Si-C}$ and $\mathrm{C-H}$
bending modes at $1200.33\:cm^{-1}$. The two frequency modes that
occur above $1300\:cm^{-1}$ are attributed to scissoring of $\mathrm{CH_{2}}$
groups. The $2200-2251\;cm^{-1}$ frequency region belongs to $\mathrm{Si-H}$
stretching with the highest intensity mode observed at $2217\:cm^{-1}$.
The $\mathrm{C-H}$ stretching modes lie in the $3100-3250\;cm^{-1}$
frequency region with the highest frequency occurring at $3128.57\:cm^{-1}$
attributed to $\mathrm{C-H}$ stretching attached to the octagon.

 There are 36 atoms in the Circular-SiC QD leading to $(3\times36)-6=102$
vibrational modes. In the calculated spectrum (Fig. \ref{fig:Raman-all}(b))
we identify three frequency regions categorized as I ($0-1150\:cm^{-1}$)
which is a broad region with several peaks, and narrow regions II
($2200-2220\;cm^{-1}$) and III ($3180-3182\;cm^{-1}$) with one high-intensity
peak each. Starting the discussion for the region I, we note that
the circular SiC lies in the $xy$ plane, where the bending modes
along positive and negative $z$-direction have been observed in the
region $0-150\:cm^{-1}$. The out-of-plane Si-H, C-H deformations
mostly occur in 0-670 cm$^{-1}$, and their in-plane bending vibrations
are in the 670-950 cm$^{-1}$ range. In the region, 150-500 $cm^{-1}$,
both in-plane and out-of-plane deformations of $\mathrm{Si-C}$, C-H,
and Si-H bonds have been observed. The frequency region $500-950\:cm^{-1}$
is attributed to the Si-H and C-H bending modes along with three $\mathrm{Si-C}$
bending modes in between. The out-of-plane deformations of Si-H and
C-H occur in the ranges $0-530\:cm^{-1}$ and $640-670\:cm^{-1}$,
while in the intermediate region the in-plane $\mathrm{Si-C}$, C-H,
Si-H bending modes are also observed at $558.04\:cm^{-1}$, $558.22\:cm^{-1}$,
and $577.47\:cm^{-1}$, respectively. The in-plane C-H and Si-H bending
modes occur in the region $670-950\:cm^{-1}$. In the frequency $950-1150\:cm^{-1}$
region, both stretching and bending modes have been observed in the
$xy$ plane, along with the C-H and Si-H bending modes. The medium
frequency region II corresponds to $\mathrm{Si-C}$ stretching modes
with the highest frequency of the entire spectrum at 2220 cm-1. Region
III is also a high-intensity region representing the $\mathrm{C-H}$
bond stretching having six normal modes.\\
Next, we discuss Tho-SiC QD with 52 atoms, and a total $(3\times52)-6=150$
modes. We note that the Raman spectrum of this QD is quite similar
to that of the circular-SiC QD in that both have a broad lower energy
region containing low-intensity peaks, followed by a couple of sharp
high-intensity peaks at higher energies. The first frequency region
varies from $0-980\:cm^{-1}$ that contains in-plane and out-of-plane
$\mathrm{Si-C}$, C--H, and Si--H bending deformations. Here, the
region from $0-480\:cm^{-1}$ is dominated by Si--C in and out of
plane deformations, whereas the region $480-980\:cm^{-1}$ has major
contributions from C--H and Si--H bending modes. The coupling of
in-plane Si--C stretching and bending modes occurs in the $980-1210\:cm^{-1}$
frequency region. The higher intensity peak at $1014.48\:cm^{-1}$
belongs to Si--H, C--H bending modes along with the Si--C stretching
and bending modes. The stretching modes of Si--H in this case begins
from $2200\;cm^{-1}$ with a total of eight modes and end at $2227.22\;cm^{-1}$.
The C--H bond stretching with 8 frequency modes lies in between $3170-3200\;cm^{-1}$.
The highest peak corresponds to the Si--H bond stretching at $2219.78\;cm^{-1}$,
and the second highest peak due to the C--H bond stretching is located
at $3180.95\;cm^{-1}$. \\
 In the case of Pho-SiC QDs, there is a total of 80 atoms leading
to $(3\times80)-6=234$ vibrational modes, while two additional bonds,
$\mathrm{C-C}$ and $\mathrm{Si-Si}$, are present in this QD as compared
to other considered QDs. Most of the Raman intensity in this QD is
confined in the energy region below 1500 cm$^{-1}$, while the higher
energy region has a few very feeble peaks. For the frequency region
$0-620\:cm^{-1}$, the bending in and out of plane deformations are
observed for each bond of QD. In between 620-700 cm$^{-1}$, the Si--Si
stretching modes are observed along with $\mathrm{Si-C}$, C--H and
Si--H bending modes, while Si--C stretching modes lie in the range
$815-1112\;cm^{-1}$. The C--C bending and C--C stretching modes
range from $1150-1200\:cm^{-1}$ and $1350-1400\:cm^{-1}$, respectively,
coupled with other bond bending deformations. Very weak Si--H intensity
has been observed in the range $2180.49-2254.05\;cm^{-1}$, with 12
Si--H stretching modes having the highest peak at $2180.88\;cm^{-1}$.
The C--H stretching modes lie in the $3196-3250\;cm^{-1}$ frequency
region with eight stretching bonds and the maximum intensity mode
at $3196.73\;cm^{-1}$. \\
For the fifth considered structure T1-SiC QD, there are 42 atoms with
a total of $(3\times42)-6=120$ vibrational modes.The $0-550\:cm^{-1}$
frequency region corresponds to in-plane ($xy$ plane) and out-of-plane
(i.e., along +ve and -ve $z$-axis) deformations of $\mathrm{C-C}$,
$\mathrm{Si-Si}$, $\mathrm{Si-C}$, $\mathrm{Si-H}$, and $\mathrm{C-H}$
bonds. The frequency region $550-600\:cm^{-1}$ contains $\mathrm{Si-Si}$
bond stretching coupled with $\mathrm{C-C}$ and $\mathrm{C-Si}$
bending modes. The region $600-850\:cm^{-1}$ dominates with $\mathrm{Si-H}$
and $\mathrm{C-H}$ bending modes and also contains a few $\mathrm{Si-C}$,
$\mathrm{C-C}$, and $\mathrm{Si-Si}$ bending deformations. The $\mathrm{C-Si}$
stretching modes lie in the frequency region $\mathrm{850-1100\:cm^{-1}}$
along with the rest of the bonds bending deformations. The region
$\mathrm{1150-1400\:cm^{-1}}$ contains C-C stretching modes coupled
with Si--H, C--H, and Si--C bending modes. The highest peak in
T1 QD belongs to this region at $\mathrm{1345.06\:cm^{-1}}$ and corresponds
to the coupled $\mathrm{C-C}$ stretching and bending modes. The 6
$\mathrm{Si-H}$ stretching modes with low intensity lie between $\mathrm{2250\;cm^{-1}-2270\;cm^{-1}}$with
maximum intensity at $\mathrm{2259.20\;cm^{-1}}$. The frequency region
for the $\mathrm{C-H}$ stretching modes ranges from $\mathrm{3120\;cm^{-1}-3150\;cm^{-1}}$
with a total of eight stretching modes, and the maximum intensity
at $\mathrm{3141.51\;cm^{-1}}$.\\
 Lastly, in this work, we have presented the Raman spectra of Triangular-SiC
QD with a total of 34 atoms, and $(3\times34)-6=96$ vibrational modes.
The first frequency starts with a low intensity region that varies
between $\mathrm{0-600\;cm^{-1}}$ containing in and out of plane
bending modes for the whole QD. The region $\mathrm{600-900\:cm^{-1}}$
corresponds to the Si--H and C--H bending modes. The frequency region
$\mathrm{900\;cm^{-1}-1050\;cm^{-1}}$ is attributed to the $\mathrm{Si-C}$
stretching modes, with the rest of the bonds undergoing bending deformations.
Here, the frequency region $\mathrm{2217.36-2235.29\;cm^{-1}}$ corresponds
to the $\mathrm{Si-H}$ stretching modes with the highest frequecy
at $\mathrm{2232.53\;cm^{-1}}$, with a total of nine $\mathrm{Si-H}$
stretching modes. All three C--H stretching modes have the same intensity
and lie in the range $\mathrm{3183.72-3183.91\;cm^{-1}}$ corresponding
to the stretching of one C-H bond at a time. 

We believe that our detailed computed Raman spectra of the considered
SiC-QDs can be used to interpret future Raman experiments on these
QDs, and help with their structural characterization\textcolor{green}{. }

\subsection{Optical Absorption Spectra}

\begin{figure}[H]
\centering{}\subfloat{\centering{}\includegraphics[scale=0.63]{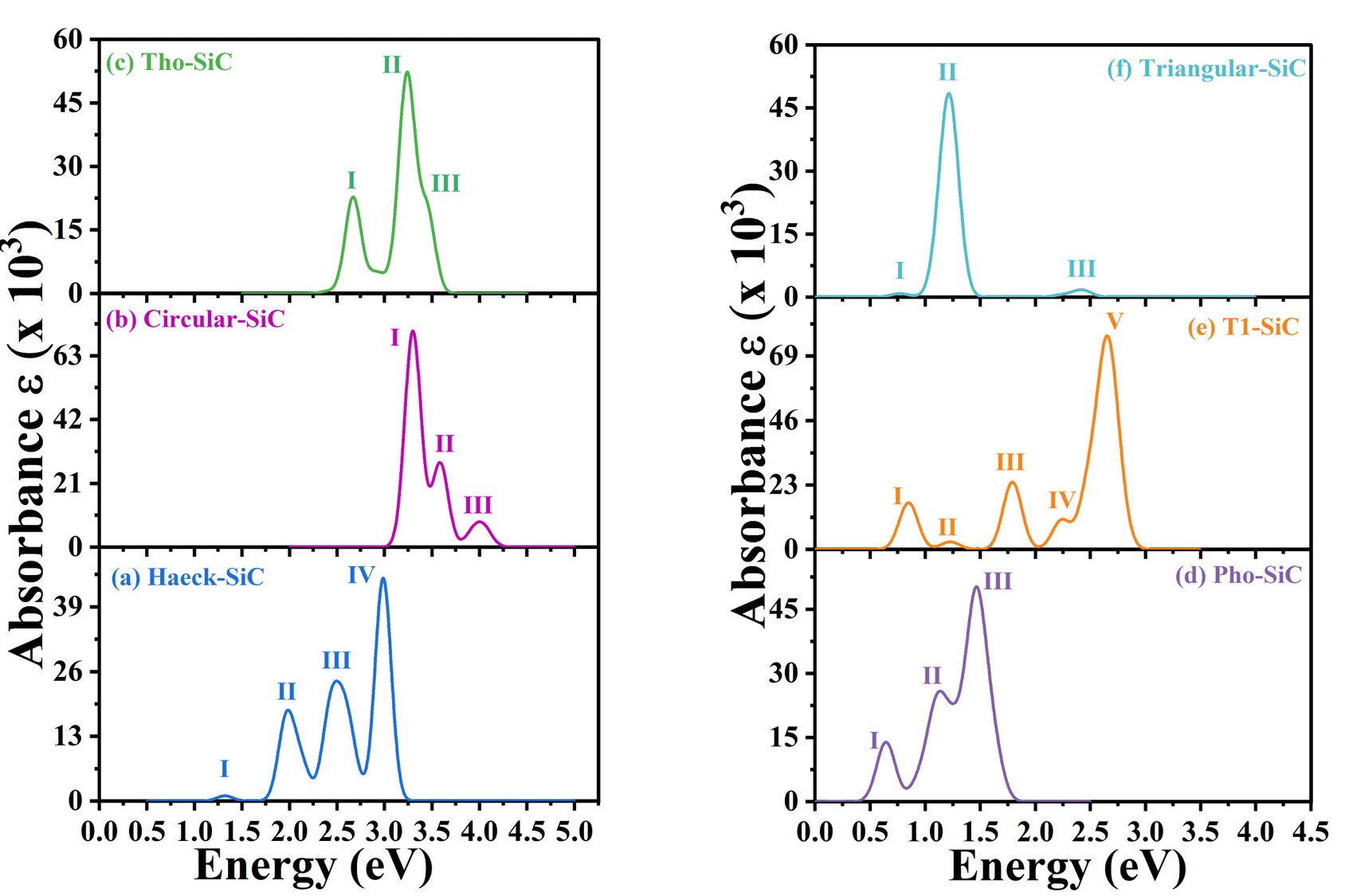}}
\caption{\label{fig:Optical-Spectra-all}Linear optical absorption spectra
for (a) Haeck-SiC, (b) Circular-SiC, (c) Tho-SiC, (d) Pho-SiC, (e)
T1-SiC, and (f) Triangular-SiC, computed at the TDDFT-B3LYP level
of theory. For plotting the spectra, a uniform line width of 0.1 eV
was assumed.}
\end{figure}

In order to compute the linear optical absorption spectra, we performed
the time-dependent-DFT (TDDFT) calculations on each QD using the B3LYP
hybrid functional and cc-pVDZ basis set, as was done for the ground-state
calculations. Absorption spectra were computed using the sum-over-states
(SOS) approach, and for this purpose 20 excited states were included
for each QD implying that the computed spectra must be quite accurate
in the UV-Vis region of the spectrum. In Fig. \ref{fig:Optical-Spectra-all}
we present the calculated TDDFT spectra of all the SiC-QDs, while
in Table \ref{tab:First-peak-energy} we present the information such
as energetic locations, wavelengths, oscillator strength (OS), dominant
configurations of their first computed excited states. From Table
\ref{tab:First-peak-energy} it is obvious that for two SiC-QDs, the
first excited states are dipole forbidden, while for several others
they are weakly allowed. Only for T1-SiC, the first peak is strongly
allowed with a significant OS. In the descending order of the first
excited state energies ($E_{1}$), the SiC-QDs can be arranged as
Circular-SiC > Tho-SiC > Haeck-SiC > T1-SiC > Pho-SiC. \textcolor{black}{The
Triangular-SiC QD is the only QD which exhibits a triplet ground state,
as a result of which all its optically excited states also have triplet
multiplicity. W}e also note that for all the QDs, $E_{1}<E_{HL}$,
because of the inclusion of the electron-hole interaction effects
in the TDDFT approach.

\textcolor{red}{{} }From the plotted absorption spectra (Fig. \ref{fig:Optical-Spectra-all})
it is obvious that all the SiC-QDs have more than one peak. Therefore,
we present detailed information about the spectra of the individual
QDs in Tables \ref{tab:Haeck-SiC}-\ref{tab:Triangular-SiC} and discuss
them next. The numbers in the parenthesis in the last column in Tables
\ref{tab:First-peak-energy}-\ref{tab:Triangular-SiC} represent the
coefficients of those configurations in the many-body wave functions
of various excited states. 
\begin{table}[H]
\centering{}\caption{\label{tab:First-peak-energy}The first excited state energy ($E_{1}$)\textcolor{blue}{,}
corresponding wavelength, $H-L$ gap ($E_{HL}$), oscillator strength
(OS), and the dominant configuration of the TDDFT wave function of
the SiC QDs. In the last column, we list the excitations contributing
to various excited states, with their coefficients in the parenthesis.}
\begin{tabular}{cccccc}
\toprule 
Structure & $E_{1}$ (eV) & $\lambda$ $(nm)$ & $E_{HL}$(eV) & OS & Dominant Configurations\tabularnewline
\midrule
\midrule 
Haeck-SiC & 1.271 & 975.54 & 2.42 & 0 & $|H\rightarrow L\rangle$ (0.7042)\tabularnewline
Circular-SiC & 2.913 & 425.63 & 3.71 & 0 & $|H-1\rightarrow L\rangle$ and c.c. (0.4283)\tabularnewline
Tho-SiC & 2.416 & 513.16 & 3.01 & 0.0039 & $|H\rightarrow L\rangle$(0.6812)\tabularnewline
Pho-SiC & 0.191 & 6495.68 & 0.66 & 0.006 & $|H\rightarrow L\rangle$(0.7232)\tabularnewline
T1-SiC & 0.848 & 1462.50 & 0.93 & 0.1235 & $\left|H\rightarrow L\right\rangle $ (0.7217)\tabularnewline
Triangular-SiC & 0.771 & 1608.21 & 1.34 & 0.0031 & $\left|L(\alpha)\rightarrow L(\alpha)+1\right\rangle $ (0.9572)\tabularnewline
\bottomrule
\end{tabular}
\end{table}

\subsubsection{Haeck-SiC}

The first transition in Haeck-SiC corresponds to a dipole-forbidden
dark state whose wave function is dominated by the excitation $H\rightarrow L$.
As shown in Fig. \ref{fig:Optical-Spectra-all}(a), the first observable
peak is feeble and located at 1.31 eV with the main contribution from
the excitation $H\rightarrow L+1$(See Table \ref{tab:Haeck-SiC}).
The next two peaks(II and III) have considerable intensity with oscillator
strengths of 0.1199 and 0.1084, respectively. The transitions $H-2\rightarrow L+1$
and $H-5\rightarrow L$ lead to a second peak corresponding to the
excitation energy of 1.97 eV. The third peak with transition energy
2.419 eV is also a linear combination of two state transitions $H\rightarrow L+3$
and $H-4\rightarrow L+1$. The fourth peak, which has the maximum
intensity, is due to an excited state located at 2.99 eV, dominated
by the transition $H-3\rightarrow L+2$. 

\begin{table}[H]
\centering{}\caption{\label{tab:Haeck-SiC}The essential information about the excited
states contributing to various peaks in the linear optical absorption
spectrum of the Haeck-SiC QD (see Fig. \ref{fig:Optical-Spectra-all}(a)).
In the wave function column, we list the excitations contributing
to various excited states, with their coefficients in the parenthesis.}
\begin{tabular}{cccc}
\toprule 
Peak & Energy(eV) & OS & Wave function\tabularnewline
\midrule
\midrule 
I & 1.311 & 0.0067 & $\left|H\rightarrow L+1\right\rangle $(0.6987)\tabularnewline
\midrule 
II & 1.966 & 0.1199 & $\begin{array}{cc}
|H-2 & \rightarrow L+1\rangle(0.6818)\\
|H-5 & \rightarrow L\rangle(0.1649)
\end{array}$\tabularnewline
\midrule 
III & 2.419 & 0.1084 & $\begin{array}{cc}
|H & \rightarrow L+3\rangle(0.6177)\\
|H-4 & \rightarrow L+1\rangle(0.1388)
\end{array}$\tabularnewline
\midrule 
IV & 2.986 & 0.3320 & |$H-3\rightarrow L+2\rangle$ (0.6193)\tabularnewline
\bottomrule
\end{tabular}
\end{table}

\subsubsection{Circular-SiC}

In Circular-SiC, which is the SiC counterpart of coronene, the lowest
excited state $E_{1}$ does not play a role in the absorption spectrum
because it is dipole forbidden (see Table \ref{tab:First-peak-energy}).
Furthermore, the second excited state whose wave function is dominated
by the excitation $H\rightarrow L$ is also dipole forbidden. The
first peak (I) is the most intense peak of the spectrum which derives
dominant contributions from $H\rightarrow L+1$ and its electron-hole
conjugate excitation $H-1\rightarrow L$, referred to as ``complex
conjugate'', and denoted as ``c.c.'', for the sake of brevity (see
table \ref{tab:Circular-SiC}). The second peak at 3.585 eV also has
a large OS with the dominant transitions as $H-1\rightarrow L+3$
and $H\rightarrow L+2$. The third peak has a comparatively smaller
OS and is dominated by transitions $H-4\rightarrow L$ and $H-5\rightarrow L+1$.

\begin{table}[H]
\centering{}\caption{\label{tab:Circular-SiC}The essential information about the excited
states contributing to various peaks in the linear optical absorption
spectrum of the circular-SiC QD (see Fig. \ref{fig:Optical-Spectra-all}(b)).
In the wave function column, we list the excitations contributing
to various excited states, with their coefficients in the parenthesis.}
\begin{tabular}{cccc}
\toprule 
Peak & Energy(eV) & OS & Wave function\tabularnewline
\midrule
\midrule 
I & 3.298 & 0.2635 & $\left|H\rightarrow L+1\right\rangle $ and c.c.(0.3469)\tabularnewline
\midrule 
II & 3.585 & 0.1021 & $\begin{array}{cc}
|H-1 & \rightarrow L+3\rangle(0.3852)\\
|H & \rightarrow L+2\rangle(-0.3839)
\end{array}$\tabularnewline
\midrule 
III & 4.043 & 0.022 & $\begin{array}{cc}
|H-4 & \rightarrow L\rangle(0.5781)\\
|H-5 & \rightarrow L+1\rangle(0.3090)
\end{array}$\tabularnewline
\bottomrule
\end{tabular}
\end{table}

\subsubsection{Tho-SiC}

In Tho-SiC, the optical gap corresponds to the $E_{1}$ state whose
wave function is dominated by the $H\rightarrow L$ excitation and
gives rise to a feeble peak in the spectrum located at 2.416 eV which
is not shown in Fig. \ref{fig:Optical-Spectra-all}(c)). The wavelength
corresponding to $E_{1}$ is 513.2 nm, which clearly lies in the visible
region (Table \ref{tab:First-peak-energy}). In Fig. \ref{fig:Optical-Spectra-all}(c)
and Table \ref{tab:Tho-SiC}, the first peak is observed at 2.669
eV, also in the visible region, which is dominated by $H-2\rightarrow L$
transition. The wave function of the most intense peak (II) located
at 3.241 eV has major contributions from $H-4\rightarrow L$ and $H-3\rightarrow L+2$
transitions. The last peak appears as a shoulder of peak II, and its
wave function is dominated by the excitation $H-1\rightarrow L+4$.

\begin{table}[H]
\centering{}\caption{\label{tab:Tho-SiC}The essential information about the excited states
contributing to various peaks in the linear optical absorption spectrum
of Tho-SiC QD (see Fig. \ref{fig:Optical-Spectra-all}(c)). In the
wave function column, we list the excitations contributing to various
excited states, with their coefficients in the parenthesis.}
\begin{tabular}{cccc}
\toprule 
Peak & Energy(eV) & OS & Wave function\tabularnewline
\midrule
\midrule 
I & 2.669 & 0.1668 & $\left|H-2\rightarrow L\right\rangle $ (0.6203)\tabularnewline
\midrule 
II & 3.241 & 0.3424 & $\begin{array}{cc}
|H-4 & \rightarrow L\rangle(0.6203)\\
|H-3 & \rightarrow L+2\rangle(0.2481)
\end{array}$\tabularnewline
\midrule 
III & 3.447 & 0.1423 & |$H-1\rightarrow L+4\rangle$(0.6305)\tabularnewline
\bottomrule
\end{tabular}
\end{table}

\subsubsection{Pho-SiC}

The optical gap of Pho-SiC is quite small at 0.191 eV, corresponding
to the $E_{1}$ state with a small OS (see Table \ref{tab:First-peak-energy})
whose wave function is dominated by the $H\rightarrow L$ excitation.
As far as its optical absorption spectrum presented in Fig. \ref{fig:Optical-Spectra-all}(d)
is concerned, the first peak at 0.644 eV involves $H-2\rightarrow L$
transition for the dominant contribution towards the wave function
along with a minor contribution from $H\rightarrow L$ excited state
(see table \ref{tab:Pho-SiC}). The second peak at 1.104 eV comprises
two excitations $H-1\rightarrow L+1$ and $H\rightarrow L+2$ with
significant contributions. The final peak is the maximum intensity
peak located at 1.447 eV with the dominant contribution from the $H\rightarrow L+4$
excitation. We note that the optical excitation energies of this QD
are much smaller as compared to others, and, could be of potential
use in low-energy optoelectronics.

\begin{table}[H]
\centering{}\caption{\label{tab:Pho-SiC}The essential information about the excited states
contributing to various peaks in the linear optical absorption spectrum
of Pho-SiC QD (see Fig. \ref{fig:Optical-Spectra-all}(d)). In the
wave function column, we list the excitations contributing to various
excited states, with their coefficients in the parenthesis.}
\begin{tabular}{cccc}
\toprule 
Peak & Energy(eV) & OS & Wave function\tabularnewline
\midrule
\midrule 
I & 0.644 & 0.1031 & $\begin{array}{cc}
|H-2 & \rightarrow L\rangle(0.6621)\\
|H & \rightarrow L\rangle(0.1462)
\end{array}$\tabularnewline
\midrule 
II & 1.104 & 0.154 & $\begin{array}{cc}
|H-1 & \rightarrow L+1\rangle(0.5232)\\
|H & \rightarrow L+2\rangle(0.3547)
\end{array}$\tabularnewline
\midrule 
III & 1.447 & 0.2234 & $\begin{array}{cc}
|H & \rightarrow L+4\rangle\end{array}$(0.6379)\tabularnewline
\bottomrule
\end{tabular}
\end{table}

\subsubsection{T1-SiC}

In Table \ref{tab:T1-SiC} we present the essential information about
the excited states contributing to the absorption spectrum of T1-SiC
(see Fig. \ref{fig:Optical-Spectra-all}(e)). The first excited state
$E_{1}$ located at 0.848 eV in T1-SiC has a significant OS in the
absorption spectrum and gives rise to peak I in the absorption spectrum.
Given the fact that its wave function is dominated by the excitation
$H\rightarrow L$, $E_{1}$ clearly corresponds to the optical gap
of the T1-SiC QD. Thus, the optical gap of this QD (0.848 eV) is less
than that of Tho-SiC (2.416 eV), but more than that of Pho-SiC (0.191
eV). The first peak is followed by a feeble peak, i.e., peak II at
1.227 eV with the dominant contributions from $H\rightarrow L+2$
and $H\rightarrow L$ excitations. The third peak (III) at 1.79 eV
is the moderately intense peak composed of two excitations $H\rightarrow L+4$
and $H-1\rightarrow L+1$. The fourth peak in the spectrum appears
as a shoulder of peak V and is characterized by the configurations
$H-1\rightarrow L+1$ and $H-1\rightarrow L+3$. The most intense
peak in the absorption spectrum (peak V) is located at 2.634 eV and
is dominated by the configurations $H-1\rightarrow L+3$ and $H-4\rightarrow L$.
Given the fact that similar to Pho-SiC, several peaks of T1-SiC QD
are also in the low-energy region, it could be useful in infrared
optics.

\begin{table}[H]
\centering{}\caption{\label{tab:T1-SiC}The essential information about the excited states
contributing to various peaks in the linear optical absorption spectrum
of T1-SiC QD (see Fig. \ref{fig:Optical-Spectra-all}(e)). In the
wave function column, we list the excitations contributing to various
excited states, with their coefficients in the parenthesis.}
\begin{tabular}{cccc}
\toprule 
Peak & Energy(eV) & OS & Wave function\tabularnewline
\midrule
\midrule 
I & 0.848 & 0.1235 & $\left|H\rightarrow L\right\rangle $ (0.7217)\tabularnewline
\midrule 
II & 1.227 & 0.02 & $\begin{array}{cc}
|H & \rightarrow L+2\rangle(0.6976)\\
|H & \rightarrow L\rangle(-0.1002)
\end{array}$\tabularnewline
\midrule 
III & 1.790 & 0.1781 & $\begin{array}{cc}
|H & \rightarrow L+4\rangle(0.6521)\\
|H-1 & \rightarrow L+1\rangle(-0.2325)
\end{array}$\tabularnewline
\midrule 
IV & 2.237 & 0.0769 & $\begin{array}{cc}
|H-1 & \rightarrow L+1\rangle(0.6344)\\
|H-1 & \rightarrow L+3\rangle(-0.219)
\end{array}$\tabularnewline
\midrule 
V & 2.634 & 0.3652 & $\begin{array}{cc}
|H-1 & \rightarrow L+3\rangle(0.6094)\\
|H-4 & \rightarrow L\rangle(0.2318)
\end{array}$\tabularnewline
\bottomrule
\end{tabular}
\end{table}

\subsubsection{Triangular-SiC}

\textcolor{black}{The spin-polarized DFT calculations on this QD predict
the ground state to have a triplet spin multiplicity, in agreement
with the well-known property of the corresponding carbon-based triangulene.
In ou}r all-electron calculations consisting of 240 electrons, we
nominally call the HOMO to be the 120th energetically ordered orbital,
irrespective of whether the orbital corresponds to up-spin ($\alpha$)
or down-spin ($\beta$). Thus the ground state has the frontier-orbital
occupancy $H^{1}(\alpha)L^{1}(\alpha)$ for the up spins, and $(H-1)^{1}(\beta)$
for the down spin. Subsequently, spin-polarized TDDFT calculations
were performed, and the computed optical absorption spectrum is presented
in Fig. \ref{fig:Optical-Spectra-all}(f), while the detailed information
about the peaks is presented in Table\textcolor{blue}{{} }\textcolor{black}{\ref{tab:Triangular-SiC}}.
We note that all the configurations contributing to the excited-state
wave functions are single excitations in the $\alpha$-spin sector,
therefore, we have suppressed the spin orientation of the excitations
in the table. Peak I of the spectrum is quite weak and corresponds
to the first excited state $E_{1}$ located at 0.771 eV, representing
the optical gap whose wave function is dominated by the excitation
$L(\alpha)\rightarrow L+1(\alpha)$. The middle peak (II) is the most
intense one located at 1.224 eV and has almost equally dominant contributions
from three excitations $L(\alpha)\rightarrow L+2(\alpha)$, $L(\alpha)\rightarrow L+3(\alpha)$,
and $H(\alpha)\rightarrow L+4(\alpha)$. The final peak is again a
very feeble one with the dominant contribution from the excitation
$H(\alpha)\rightarrow L+8(\alpha)$.
\begin{table}[H]
\centering{}%
\begin{tabular}{cccc}
\toprule 
Peak & Energy(eV) & OS & Wave function\tabularnewline
\midrule
\midrule 
I & 0.771 & 0.0031 & $\left|L\rightarrow L+1\right\rangle $(0.9572)\tabularnewline
\midrule 
II & 1.224 & 0.1541 & $\begin{array}{cc}
|L & \rightarrow L+2\rangle(0.5446)\\
|L & \rightarrow L+3\rangle(0.5062)\\
|H & \rightarrow L+4\rangle(0.5033)
\end{array}$\tabularnewline
\midrule 
III & 2.424 & 0.0063 & $\begin{array}{cc}
|H & \rightarrow L+8\rangle(0.9317)\\
|L & \rightarrow L+5\rangle(0.2063)
\end{array}$\tabularnewline
\bottomrule
\end{tabular}\caption{\label{tab:Triangular-SiC}The essential information about the excited
states contributing to various peaks in the linear optical absorption
spectrum of Triangular-SiC QD (see Fig. \ref{fig:Optical-Spectra-all}(f)).
In the wave function column, we list the excitations contributing
to various excited states, with their coefficients in the parenthesis.
For this QD, all the excitations are in the $\alpha$-spin sector.}
\end{table}
 From Table \ref{tab:Triangular-SiC}, we note that all the peaks
lie in the range spanning the infrared to the visible region of the
optical spectrum. 

\section{Conclusions and Future Directions}

In this work, we studied the structural, electronic, vibrational,
and optical properties of six nanometer-sized finite fragments, i.e.,
quantum dots, of several recently proposed 2D monolayers of SiC. The
considered QDs have different topological characteristics, with their
edges passivated by hydrogen atoms. The binding energy and vibrational
frequency calculations indicated that all six predicted SiC-QDs are
thermodynamically and structurally stable. The computed Raman as well
as optical absorption spectra contain structural fingerprints of different
QDs, thus, we believe that once synthesized, these spectroscopies
can be used to identify them.

The optical gaps of the considered QDs show significant variation
from 0.64 eV (Pho-SiC) to 3.30 eV (Circular-SiC), indicating that
our predicted SiC-QDs can be useful in devices operating from the
infrared to the ultraviolet region of the optical spectrum. 

On examining the TDDFT many-particle wave functions of the optically
active excited states of these QDs, we found the dominant contributing
configurations to have $\pi\rightarrow\pi^{*}$ character, with negligible
contributions from excitations involving $\sigma/\sigma*$ orbitals.
This is similar to the case of planar aromatic hydrocarbons for which
$\sigma$-$\pi$ separation holds good when it comes to low-lying
optically excited states in the UV-Vis region. Therefore, one can
obtain an effective $\pi$-electron Hamiltonian for planar SiC-QDs
for describing their optical excitations, similar to the PPP model\citep{soos1984valence}
for the aromatic hydrocarbon molecules and H-passivated graphene fragments
used extensively in our group\citep{rai-scientific-reports,basak2015theory,basak2021graphene}.
We plan to develop such an effective model Hamiltonian for SiC and
other similar planar $\pi$-conjugated systems in future because it
will allow us to perform more rigorous electron-correlated calculations
due to a drastic reduction in the degrees of freedom achieved in making
a transition from the first-principles all-electron approach to the
one involving only the $\pi$-electrons.

Furthermore, this work was restricted only to the study of the linear
optical response of these SiC-QDs. However, $\pi$-conjugated molecules,
particularly aromatic hydrocarbons are known for their strong nonlinear
optical response (NLOR) as well. However, given their high level of
symmetry, the first NLOR in aromatic molecules is typically in the
third order. On the other hand, for the heteroatom $\pi$-conjugated
structures such as SiC-QDs considered here, we expect their NLOR to
be in the second order, therefore, it will be worthwhile to study
processes such as second-harmonic generation in them. At present calculations
along these lines are underway in our group, and the results will
be presented in future publications.

\section*{Author Information }

\subsection*{Corresponding Authors}

Alok Shukla:  {*}E-mail: shukla@phy.iitb.ac.in

\subsection*{Notes}

The authors declare no competing financial interests.

\section*{Acknowledgements}

One of the authors, RJ acknowledges financial assistance from CSIR
JRF. 

\bibliographystyle{achemso}
\bibliography{SiC}

\end{document}